\documentclass[aps,pra,reprint,superscriptaddress]{revtex4-2}

\usepackage{mathtools,amssymb,amsmath,commath}
\usepackage{graphicx}
\usepackage{csquotes}

\usepackage{xcolor}

\DeclareMathOperator{\sech}{sech}
\DeclareMathOperator{\sgn}{sgn}

\usepackage[colorlinks=true,bookmarks=true,allcolors=blue,citecolor=red,urlcolor=blue]{hyperref}

\begin{document}
	\title{Comprehensive study on beam dynamics inside symmetrically chirped waveguide array mimicking the graded index media }
	\author{Anuj P. Lara}
	\affiliation{Department of Physics, Indian Institute of Technology Kharagpur, Kharagpur 721302, India}
	
	\author{Samudra Roy}
	\email{samudra.roy@phy.iitkgp.ac.in}
	\affiliation{Department of Physics, Indian Institute of Technology Kharagpur, Kharagpur 721302, India}
	
	\begin{abstract}
		 In this article, we explore the beam dynamics within symmetrically chirped nonlinear waveguide arrays, focusing on linear and quadratic chirping schemes. We propose a practical structure for these arrays that enhances control over light propagation. By employing a continuous approximation of the discrete nonlinear Schrödinger equation (DNLSE), we utilize a semi-analytical variational method to analyze beam behavior under waveguide chirping. Our findings indicate that the symmetrically chirped waveguide arrays behave similarly to graded index systems, with varying coupling coefficients analogous to the refractive index in continuous media. We derive a steady-state solution and validate it against numerical simulations, alongside conducting a linear stability analysis to assess the robustness of these solutions. The results reveal that input Gaussian beams in such waveguide arrays follow an oscillatory trajectory akin to that in parabolic index media. Notably, under nonlinear conditions, these beams  evolve as discrete solitons. Our rigorous investigation of the propagation characteristics in both linear and nonlinear regimes highlights the intricate dynamics of optical beams within the engineered chirped waveguide arrays, supported by comparisons to comprehensive numerical simulations.

	\end{abstract}
	
	\maketitle
	
	\section{Introduction}
	
    Waveguide arrays (WAs) have emerged as a valuable photonic platform, enabling a wide range of applications in optical communications and fundamental physics. These arrays, characterized by their periodic arrangement of waveguide channels, facilitate unique optical dynamics on a macroscopic scale, reflecting behaviors previously observed only at atomic and molecular levels. Notable phenomena such as Bloch oscillations \cite{pertsch_optical_1999,morandotti_experimental_1999}, Peierls-Nabarro potential\cite{kivshar_peierls-nabarro_1993}, and Anderson localization\cite{lahini_anderson_2008}, which are fundamental to crystal lattices, have been demonstrated in WAs. Additionally, WAs have been linked to various relativistic phenomena, including Klein tunneling \cite{longhi_klein_2010}, \textit{Zitterbewengung} \cite{dreisow_classical_2010,longhi_photonic_2010} and neutrino oscillation \cite{marini_optical_2014}, as well as classical analogues of quantum coherence displaced Fock states \cite{perez-leija_glauberfock_2010,keil_classical_2011}, Dirac solitons \cite{tran_dirac_2015}, etc. The research also highlights the formation of stable higher-dimensional solitons and light bullets \cite{mihalache_spatiotemporal_2008,minardi_three-dimensional_2010,vinayagam_stable_2018}, surface \cite{suntsov_observation_2006} and vortex solitons \cite{dong_vortex_2022}, along with successful demonstrations of beam routing, steering \cite{rosberg_demonstration_2006,verslegers_deep-subwavelength_2009,li_self-deflecting_2015,heck_highly_2017}, and lensing \cite{catrysse_complete_2015} within modulated photonic arrangements. Recently, WAs have attracted interest due to their potential applications in parity-time (PT) symmetry \cite{wimmer_observation_2015,kalozoumis_pt_2016,wu_floquet_2021}, topology, \cite{el-ganainy_non-hermitian_2018,kim_recent_2020},  multiplexing \cite{cerutti_engineering_2017}, demultiplexing \cite{benisty_transverse_2015}, synthetic absorbers \cite{teimourpour_non-hermitian_2017} and lasing \cite{parto_edge-mode_2018}. Further advances have been made for programmable quantum circuits \cite{yang_programmable_2025} and entangled photon generation and distribution \cite{solntsev_spontaneous_2012,matsuda_generation_2017,raymond_tunable_2024}.
    
Various applications of WAs involve the introduction of both ordered and disordered perturbations into their periodic structures. Disordered modulations can lead to optical phenomena such as Anderson localization \cite{lahini_anderson_2008,cheng_observation_2022}, while ordered perturbations facilitate topological effects and non-Hermitian dynamics \cite{el-ganainy_non-hermitian_2018,wu_floquet_2021}, as well as phenomena like \textit{Zitterbewegung} \cite{dreisow_classical_2010}. A specific form of ordered modulation, known as chirping, can be implemented via electro-optic \cite{peschel_optical_1998} and thermo-optic \cite{pertsch_optical_1999} effects, introducing a transverse gradient in the propagation constant. Additionally, chirping can also be achieved by altering geometry through femto-second laser writing \cite{rosberg_demonstration_2006}  or fragmenting the waveguide structure \cite{Belabas2009}. This modulation technique has demonstrated efficacy in applications such as beam routing \cite{li_self-deflecting_2015,lara_dynamic_2020}  and lensing \cite{catrysse_complete_2015,honari-latifpour_arrayed_2022}, mimicking a parabolic index profile.

This article examines the beam dynamics within symmetrically chirped one-dimensional nonlinear WAs, focusing on two chirping schemes: \textit{linear} and \textit{quadratic}. The study establishes a framework for determining the functional forms of the coupling coefficient, which varies in the transverse direction. Using this framework, we derive an approximate continuous version of the discrete coupled mode equations, leading to the continuous paraxial wave equation and the \textit{nonlinear Schrödinger equation} (NLSE) under different power regimes. The derivation highlights that additional terms arising from the chirping act as an effective potential, affecting diffraction characteristics. By applying a semi-analytical variational technique to the 1D paraxial wave equations, the study produces expressions for the evolution of Gaussian beams corresponding to both chirping schemes, drawing parallels with beam evolution in continuous graded index media. We analyze stationary solutions and self-focusing beam filamentation within the discrete system and extend their investigation into higher power scenarios to capture discrete soliton dynamics. The theoretical predictions of oscillatory soliton behavior in symmetrically chirped WAs are validated against numerical simulations, revealing distinctive characteristics of beam dynamics not previously explored in engineered waveguide configurations.

	\section{Theory}

To examine the beam dynamics within a WA, we analyze a semi-infinite series of identical nonlinear waveguides that are weakly coupled to their nearest neighbors through evanescent coupling. These waveguides are considered ideal, free from any deformations or losses. The dynamics of mode amplitude evolution in the \( n^{\text{th}} \) waveguide, under continuous-wave excitation and nearest-neighbor coupling, is governed by the \textit{discrete nonlinear Schr\"{o}dinger equation} (DNLSE) \cite{kivshar_peierls-nabarro_1993,cai_localized_1994,christodoulides_discrete_1988,eisenberg_discrete_1998,morandotti_self-focusing_2001}:
	\begin{align}
		i \frac{d E_n}{dz}  + C_{n}^{n+1} E_{n+1} + C_{n-1}^{n} E_{n-1} + \gamma \abs{E_n}^2 E_n = 0.
		\label{eq:dnlse}
	\end{align}
	Here $E_n$ is the electric field amplitude in the $n^\text{th}$ waveguide, $C_n^{n+1}$ and $C_{n-1}^n$ represent the coupling coefficients of the $n^{\text{th}}$ waveguide to its nearest neighbors $(n+1)^\text{th}$ and $(n-1)^\text{th}$ waveguides, respectively, and $\gamma$ being the nonlinear coefficient arising due to Kerr effect.
	For an analytical study, it is beneficial to use a normalized form of Eq. \eqref{eq:dnlse} by making the transformations,  $\eta_n^{n+1} \rightarrow C_n^{n+1}/C_0^1$, $\xi \rightarrow C_0^1 z$, $E_n \rightarrow \sqrt{P_0} a_n$, and $\mathcal{N}^2 \rightarrow \gamma P_0/C_0^1$ to obtain,
	\begin{align}
		i \frac{d a_n}{d \xi} + \eta_n^{n+1} a_{n+1} + \eta_{n-1}^n a_{n-1} + \mathcal{N}^2\abs{a_n}^2 a_n = 0.
		\label{eq:norm_dnlse}
	\end{align}
	Here, $P_0$ is the peak beam power, and $C_0^1$ is the coupling coefficient between the $n=0$ and $n=1$ waveguides that acts as a reference coupling coefficient in absence of any chirping.
	$\mathcal{N}$ is equivalent to the soliton order, where $L_{NL} = 1/\gamma P_0$ is the nonlinear length \cite{agrawal_nonlinear_2013}.
	For a uniform waveguide array with identical coupling coefficients $\eta_{n-1}^{n} = \eta_n^{n+1} = \eta$, Eq. \eqref{eq:norm_dnlse} reduces to 
	\begin{align}
		i \frac{d a_n}{d \xi}  + \eta (a_{n+1} + a_{n-1}) + \mathcal{N}^2 \abs{a_{n}}^2 a_n = 0,
		\label{eq:uniform_wa}
	\end{align}
	exhibiting a $\sech$ soliton-like solution for $\mathcal{N} = 1$ \cite{christodoulides_discrete_1988,eisenberg_discrete_1998}.
In a linear waveguide, under low power conditions, the dispersion relation links the propagation constant ($\beta$) and the transverse wavenumbers ($\kappa$) for a discrete plane wave defined by $a_n(\xi) = a_0 \exp \left[i (n \kappa + \beta \xi)\right]$, resulting in $\beta(\xi) = 2 \eta \cos(\kappa)$. This relation leads to the formulation of the general solution for a propagating discrete soliton as \cite{tran_diffractive_2013}
	\begin{align}
		a_{\text{sol}}(\xi,n) = a_0 \sech \left[\frac{n a_0}{\sqrt{2 \eta \cos(\kappa_0)}}\right] e^{i a_0^2 \xi/2},
		\label{eq:sechsol}
	\end{align}
	where $a_0$ and $\kappa_0$ are the initial values of the soliton amplitude and central transverse wavenumber, respectively.

	\subsection{Waveguide array design}
	
   To create a system capable of supporting a discrete soliton, we can employ femto-second laser waveguide writing facility \cite{Szameit2006,lv_three-dimensional_2015,pavlov_femtosecond_2017,wang_direct_2019,wang_two-photon_2023}. We propose an index array of 1.8\% GeO$_2$-doped silica cores within a silica cladding. At an operating wavelength of $\lambda_0 = 1.55$ $\mu$m, the refractive indices are determined to be $\mu_1 = 1.4477$ for the core and $\mu_2 = 1.4446$ for the cladding \cite{fleming_dispersion_1984}. The cylindrical cores, with a radius of $R = 6$ $\mu$m, have a Kerr coefficient of $n_2 \approx 2.7 \times 10^{-20} ~\text{m}^2 W^{-1}$, resulting in a nonlinear coefficient for the fundamental mode of $\gamma = 0.64$ W$^{-1}$ km$^{-1}$. For such an array of cylindrical waveguides, the coupling coefficient between adjacent channels is given as \cite{tewari_analysis_1986}
	\begin{align}
		\label{eq:coup_coeff}
		C = \frac{\lambda_0}{2 \pi \mu_1} \frac{U^2}{R^2 V^2} \frac{K_0 (W d /R)}{K_1^2(W)}.
	\end{align}
	Here, $d$ is the separation between the two waveguide channels; $\lambda_0$ is the free space wavelength ($1.55 ~\mu$m in this case); $R$ is the core radius; $K_{j}$ are the modified Bessel functions of the second kind of order $j$.
	$U$ and $W$ are the transverse mode parameters that satisfy $U^2 + W^2 = V^2$, where the $V$ parameter is defined as $V = ({2 \pi R} \sqrt{\mu_1^2 - \mu_2^2})/{\lambda_0}$.
	$U$ is approximated as $U \approxeq 2.405 e^{-(1 -v/2)/V}$, with $v = 1 - (\mu_2/\mu_1)^2$ \cite{snyder_coupled-mode_1972}.
	For a reference separation of $d = 30 ~\mu \text{m}$  we obtain a coupling coefficient of $C = 3.26~ \text{m}^{-1}$ that can support a fundamental discrete soliton ($\mathcal{N}^2=1$) with a peak beam power of $P_0 = C/\gamma \approx 5.6 ~\text{kW}$.
	
	A chirped WA can be achieved through modulation of the material properties of individual waveguides using electro/thermo-optic effects\cite{peschel_optical_1998,pertsch_optical_1999,yang_programmable_2024} or by altering the geometry of the waveguides \cite{eisenberg_diffraction_2000,cao_asymmetric_2013,li_self-deflecting_2015,catrysse_complete_2015,wang_direct_2019}. The coupling coefficient is affected by the inter-waveguide separation ($d$), while other parameters, such as the propagation constant and nonlinear coefficients, remain constant.
	{
		Such variations in the geometry can be achieved by femto-second laser writing, which provides a high control over the waveguide dimensions ($<1~\mu \text{m}$) and position in the range of a few nanometers ($\sim 32~\text{nm}$) \cite{wang_precise_2024,winkler_femtosecond_2025}.
		In addition, UV and electron beam lithography can also be utilized to create such arrays with ridge/rib waveguides which offer high precision in the waveguide geometry \cite{yamada_single_2017,zheng_uv-led_2021,wang_optofluidic_2024}.	
	
	}

	\subsubsection{Linear Chirping}
	
		\begin{figure}[h!]
		\centering
		\includegraphics[width=\linewidth]{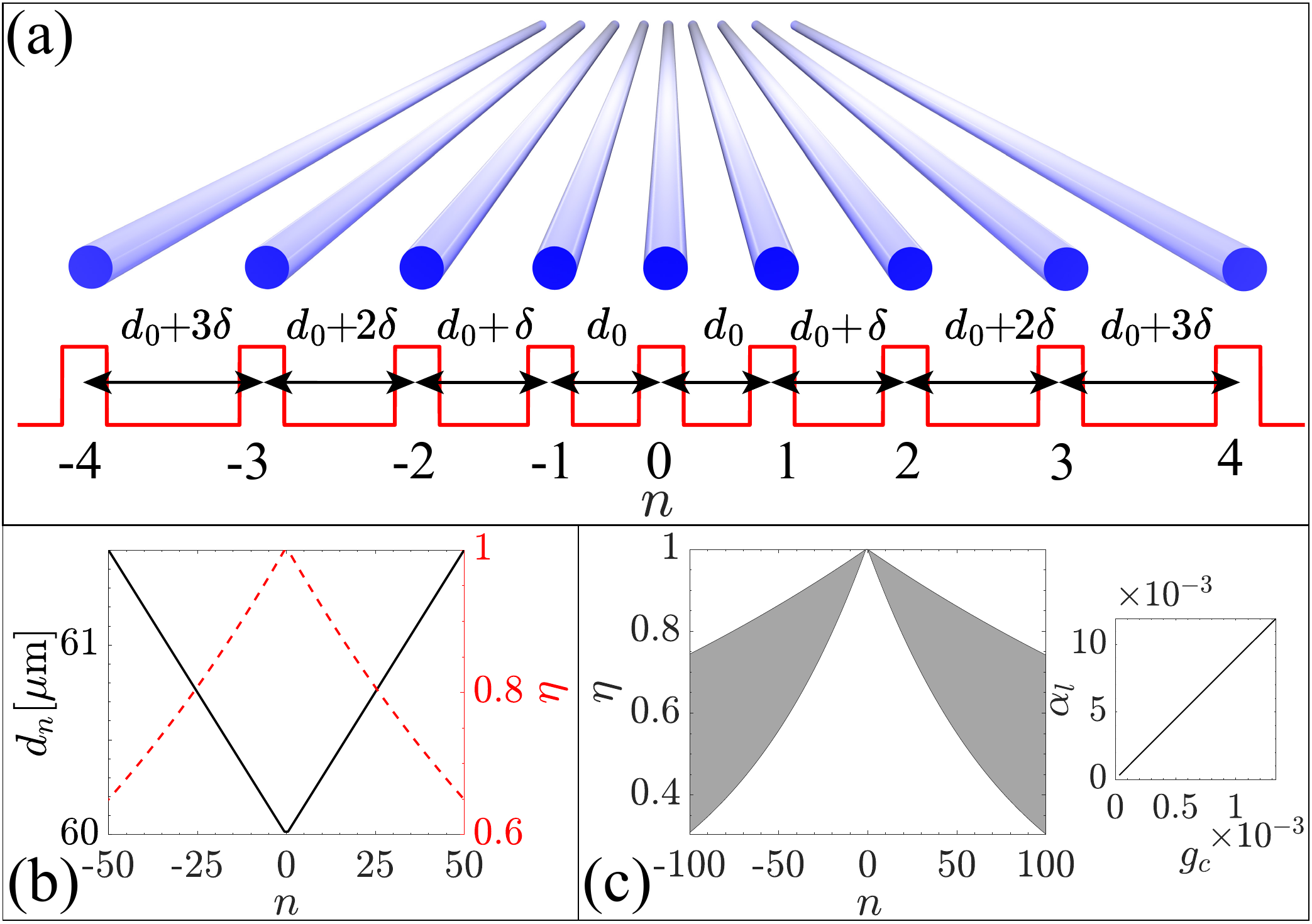}
		\caption{(a) Schematic representation of the waveguide arrangement under a symmetric linear chirp, along with the variation in refractive index in the transverse direction.
			(b) The variation in the inter-waveguide separation $d_n$ (solid black line) and the corresponding normalized coupling coefficient $\eta$ (dashed red line) as a function of $n$, for  typical parameters $\delta = 30~\text{nm}$ and  $d_{\text{ref}} = 30~\mu \text{m}$.
			(c) Variation of the coupling coefficient $\eta$ along the transverse coordinate $n$
			for the chirping strength $\delta$ ranging from 10 nm to 40 nm (shaded region). In the right inset we plot $\alpha_l$ as a function of $g_c$.
		}
		\label{fig:linear_scheme}
	\end{figure}

A linearly chirped waveguide array (WA) features a systematic increase in the spacing between adjacent waveguides in the transverse direction, modeled as a function of the waveguide index ($n$). The separation between the $(n+1)^\text{th}$ and $n^\text{th}$ waveguides, denoted as $d_n$, is symmetric about the central waveguide ($n=0$), following the relationship $d_n = d_{-(n+1)}$. This arrangement can be described mathematically as $d_n = d_{\text{ref}} + \delta\left|n + 1/2\right|$, where $d_{\text{ref}}$ is the reference separation in the absence of chirping, and $\delta$ is denoted as the chirping strength. A schematic representation (see \figref{fig:linear_scheme}) of this arrangement illustrates the varying refractive index and waveguide separations across an array of 101 waveguides, utilizing typical parameters such as $\delta = 30~\text{nm}$ and $d_{\text{ref}} = 30~\mu \text{m}$. Furthermore, the dependence of coupling coefficients on different chirping strengths is highlighted, with specified boundaries for $\delta = 10~\text{nm}$ and $\delta = 40~\text{nm}$.
The exponential decay of the coupling coefficient with respect to the transverse variable \( n \) is explained by the decaying nature of the \( K_0 \) function, which aligns with experimental findings \cite{Szameit2006}. For the parameters considered, the modified Bessel function can be approximated as \( K_0(x) \approx \sqrt{\frac{\pi}{2 x}} e^{-x} \), where \( x = Wd_n/R \). Assuming a weak chirping value (\( \delta \ll d_0 \)), the variation in the denominator is minimal relative to the numerator, leading to an approximation for the coupling coefficient as \( C_n^{n+1} \approx C_0^1 e^{-\alpha_l \abs{n + 1/2}} \), with \( \alpha_l = mg_c \), where \( m = W d_\text{ref}/R \) and \( g_c = \delta/d_\text{ref} \) represents normalized chirping strength. The normalized coupling coefficient variation can be modeled as \( \eta_n^{n+1} = C_n^{n+1}/C_0^1 \), fitting the function \( e^{-\alpha_l \abs{n + 1/2}} \) for extraction of \( \alpha_l \). It is noted that \( \alpha_l \) varies linearly with \( g_c \), exhibiting a slope of \( m = 8.7 \), consistent with \( m = W d_0/R \approx 8.6 \).

	\subsubsection{Quadratic Chirping}

\begin{figure}[t!]
		\centering
		\includegraphics[width=\linewidth]{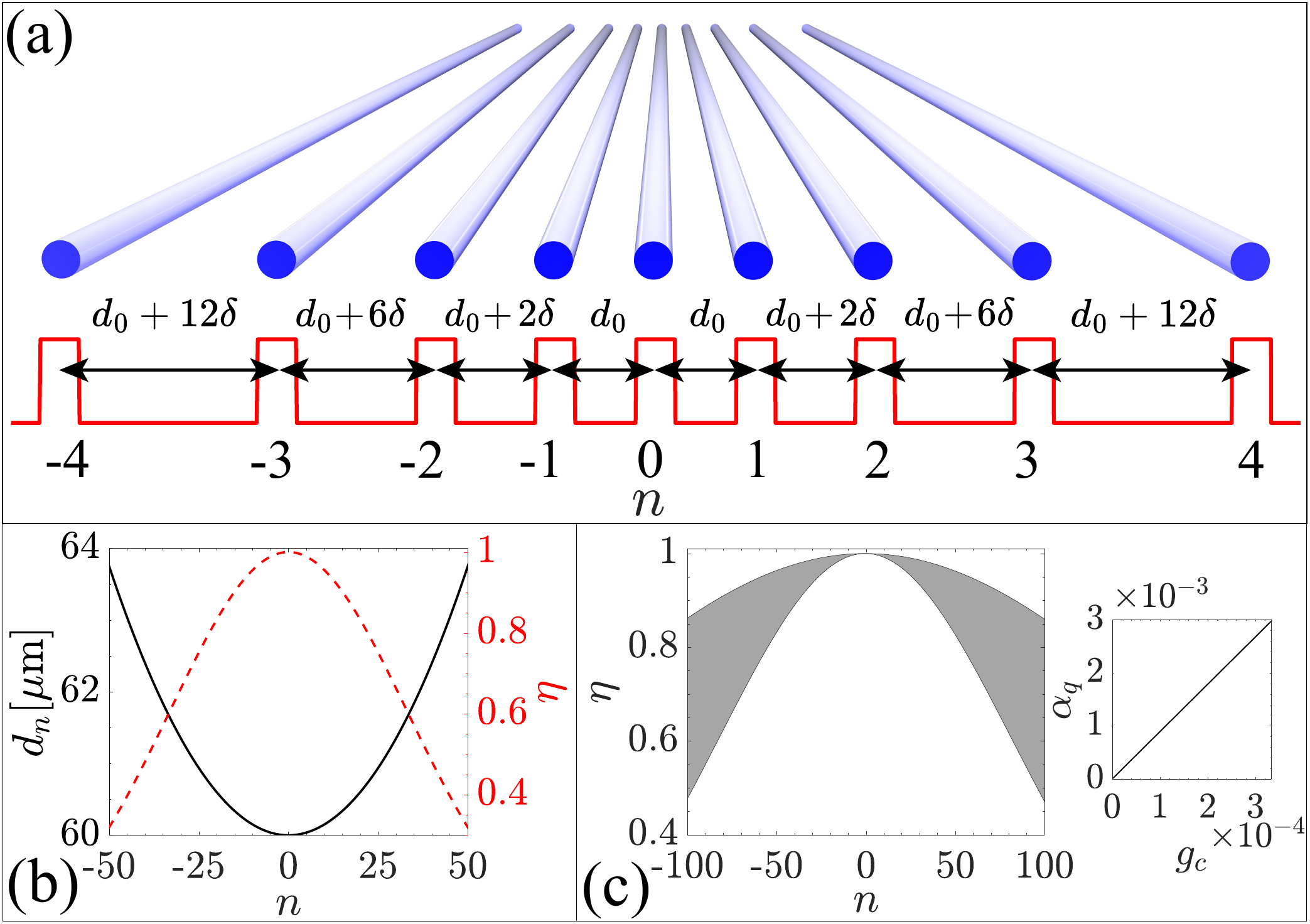}
		\caption{(a) Schematic representation of the waveguide arrangement under a symmetric quadratic chirp, along with the variation in refractive index in the transverse direction.
			(b) The variation of $d_n$, the separation between adjacent waveguides, (solid black line) and the corresponding normalized coupling coefficient profile $\eta$ (dashed red line) for $\delta = 1.5$ nm along the transverse coordinate $n$.
			(c) Variation of the coupling coefficient $\eta$ along the transverse coordinate $n$
			for the chirping strength $\delta$ ranging from 0.05 nm to 0.25 nm (shaded region). In the right inset we plot $\alpha_q$ as a function of $g_c$.
		}
		\label{fig:quad_scheme}
	\end{figure}

The symmetric quadratic chirping scheme involves a waveguide channel separation that varies quadratically in the transverse direction, defined by the equation \(d_n = d_{\text{ref}} + \delta (n + 1/2)^2\). This setup is illustrated in \figref{fig:quad_scheme} (a), which includes the transverse refractive index profile and the separation variation as a function of the waveguide index \(n\). The coupling coefficient, illustrated in the same figure (see plot (b)), is calculated for a reference separation of \(d_{\text{ref}} = 30 \mu \text{m}\) and varies with different strengths of chirping, denoted by \(\delta\). An approximate relationship for the coupling coefficient along \(n\) is expressed as \(\eta_n^{n+1} \approx e^{-\alpha_q(n + 1/2)^2}\), where \(\alpha_q\) signifies the quadratic chirping coefficient. Numerical results align closely with the approximate formulations, revealing a linear correlation between \(\alpha_q\) and the normalized chirping strength \(g_c\). Consequently, a clear functional expression for the coupling coefficient \(\eta(n)\) is established for both chirping schemes, which will be utilized in the subsequent section to derive the continuous variant of the DNLSE.

	\subsection{Continuous approximation of the DNLSEs}

In this work, we propose an analytical approach utilizing the variational technique \cite{anderson_pereira-stenflo_1999} to approximate a solution for the DNLSE Eq. \eqref{eq:norm_dnlse}. The discrete aspects of the WAs present a challenge for analytical treatment, which can be addressed by adopting a continuous approximation for the transverse variable $n$. This approximation is valid particularly when the analyzed beams involve a significant number of waveguides, typically exceeding five \cite{christodoulides_discrete_1988,tran_diffractive_2013}. By treating the mode distribution as a continuous function, we transform the nearest neighbor mode through a Taylor expansion \cite{christodoulides_discrete_1988}
	\begin{align}
		\psi(n \pm 1, \xi) = \psi(n,\xi) \pm \partial_n \psi(n,\xi) + \frac{1}{2}\partial_n^2 \psi(n,\xi).
	\label{eq:cont_approx}	
	\end{align}
	Incorporating these modifications in Eq. \eqref{eq:norm_dnlse}, we obtain the {quasi-}continuous counterpart  of the DNLSE as,
	\begin{align}
		i \partial_{\xi} \psi +\abs{\psi}^2 \psi + \left(\eta_n^{n+1} + \eta_{n-1}^n \right) \left[\psi + \frac{1}{2}\partial_n^2 \psi\right] \nonumber \\
		+ \left( \eta_n^{n+1} - \eta_{n-1}^n \right) \partial_n\psi = 0.
		\label{eq:cont_v1}
	\end{align}

\begin{figure}[h!]
		\centering
		\includegraphics[width=\linewidth]{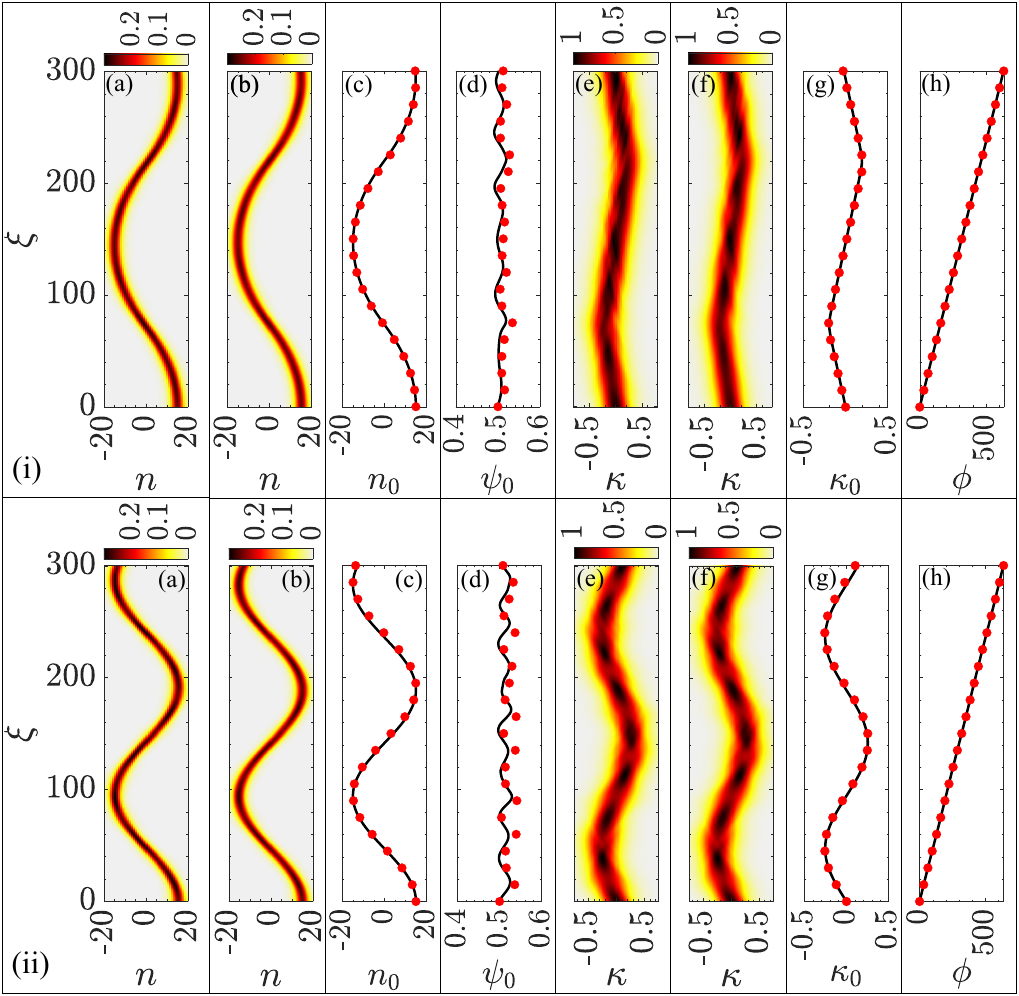}
		\caption{Comparison of a soliton evolving as per the DNLSE and the approximate NLSE under (i) linear and (ii) quadratic chirping.
		(a) and (b) are the evolution in $n$ space along with the corresponding evolution in the wavenumber ($\kappa$) space (e) and (f) as per the DNLSE and NLSE, respectively.
		The evolution of the parameters as per the DNLSE (solid red dots) and NLSE (solid black line) are given in (c) $n_0$, (d) $\psi_0$, (g) $\kappa_0$, and (h) $\phi$.
		}
		\label{fig:nlse_dnlse_comp}
	\end{figure}

The waveguide chirping, manifested in the coupling coefficients $\eta_n^{n+1}$ and $\eta_{n-1}^n$, can be linear or quadratic. 
For linear chirping, $\eta_n^{n+1} = e^{-\alpha_l |n + 1/2|}$ leads to a modified 
{ NLSE}, especially for small $\alpha_l$ ($\approx 10^{-3}$), as (for details see Appendix \ref{appen1}), 
	\begin{align}
		i\partial_{\xi} \psi + \partial_n^2 \psi + 2 \psi + \mathcal{N}^2\abs{\psi}^2 \psi = &\alpha_l \abs{n}\left(2 \psi + \partial_n^2 \psi\right) \nonumber \\
		& + \sgn(n) \alpha_l \partial_n \psi.
		\label{eq:cont_linear_2}
	\end{align} 
In contrast, quadratic chirping, represented by $\eta_n^{n+1} = e^{-\alpha_q (n + 1/2)^2}$, yields a different effective { NLSE} as (for details see Appendix \ref{appen1}), 
\begin{align}
		i\partial_\xi \psi + 2\psi + \partial_n^2 \psi + \mathcal{N}^2\abs{\psi}^2 \psi = &\alpha_q n^2 \left(2\psi + \partial_n^2 \psi\right) \nonumber \\
		&+ 2 \alpha_q n \partial_n \psi.
		\label{eq:cont_quad_2}
	\end{align}
The right-hand side of both Eq.\eqref{eq:cont_linear_2} and \eqref{eq:cont_quad_2} reflects effects of chirping, including an effective potential, a gradient in the diffraction coefficient, and transverse motion due to asymmetry in the coupling. The potentials exhibit triangular and parabolic shapes for linear and quadratic chirping, respectively. Numerical simulations validate the derived equations, employing discrete solitons $a_n = \psi_0 \sech \left[\frac{\psi_0 (n - n_0)}{\sqrt{2 \eta \cos(\kappa_0)}}\right]e^{i[\phi + \kappa_0(n-n_0)]},$ with the initial values of $\psi_0(0) = 0.5$, $n_0(0) = 15$, $\kappa_0(0) = 0$, and $\phi(0) = 0$. Results from these simulations are documented in \figref{fig:nlse_dnlse_comp}.
In the simulation, the strengths of linear and quadratic chirping were set at 10 nm and 0.5 nm, respectively. The numerical solutions of the fundamental DNLSE and its continuous approximations exhibited strong agreement, affirming the accuracy of the derived governing equations. Additionally, different soliton parameters identified through numerical methods were consistent across both equations. This validation not only confirms the reliability of the governing  Eq.\eqref{eq:cont_linear_2} and Eq.\eqref{eq:cont_quad_2} but also establishes a solid foundation for employing variational methods to explore the intricate dynamics of solitons in a chirped WA.

   \section{Variational Analysis and Results}
	
   In the preceding section, we derived a continuous version of the fundamental DNLSE to facilitate the application of Variational Analysis (VA) in the chirped waveguide model. Our investigation is divided into two domains based on beam power: $(a)$ the linear domain, where nonlinear effects are disregarded by omitting the Self-Phase Modulation (SPM) term, allowing us to analyze the evolution of a Gaussian beam under two chirping schemes; and $(b)$ the nonlinear domain, where we seek soliton-like solutions characterized by a \(sech\) form for uniform WA \cite{tran_diffractive_2013}. Utilizing a \(sech\) \textit{ansatz}, we apply VA to elucidate the dynamics of the propagating solitonic beam while considering waveguide chirping as a perturbative factor.

	\subsection{Linear Case (in absence of nonlinearity)}
	In the regime of low powers, the SPM term can be neglected ($\mathcal{N}^2 \rightarrow 0 $) and for the linearly chirped WA the NLSEs Eq. \eqref{eq:cont_linear_2}  is written as,  
	\begin{align}
		i\partial_{\xi} \psi + \partial_n^2 \psi + 2 \psi  -\alpha_l \abs{n}\left(2 \psi + \partial_n^2 \psi\right) - \sgn(n) \alpha_l \partial_n \psi = 0.
		\label{eq:lin_lin}
	\end{align}
	Similarly for the case of quadratic chirping Eq.\eqref{eq:cont_quad_2} can be rearranged as,
	\begin{align}
		i\partial_\xi \psi + \partial_n^2 \psi + 2\psi - \alpha_q n^2 \left(2\psi + \partial_n^2 \psi\right) - 2 \alpha_q n \partial_n \psi = 0.
		\label{eq:lin_quad}
	\end{align}
	 To initiate the variational analysis, we first construct the Lagrangian density, \(\mathcal{L}\), for the governing equations. \(\mathcal{L}\) is reduced $L =\int_{-\infty}^{\infty} \mathcal{L} dn$ by appropriate \textit{ansatz}.
     Utilizing the \textit{Rayleigh-Ritz optimization method} \cite{anderson_pereira-stenflo_1999}, we derive the \textit{Euler-Lagrange equation}, $\partial_{\xi} (\partial_{\nu_{\xi}} L) - \partial_\nu L=0$, which describes the dynamics of the beam through a set of coupled ordinary differential equations (ODEs). For the linear case, we adopt a Gaussian beam distribution as our \textit{ansatz}, represented explicitly as, 
     \begin{align}
		\psi = \psi_0 \exp\left[ -\frac{n^2}{2 n_w^2} + i \phi + i \theta n^2 \right],
		\label{eq:gauss_ansatz}
	\end{align}
     where the key beam parameters,   amplitude ($\psi_0$), width ($n_w$), phase ($\phi$), and phase front curvature ($\theta$) are allowed to vary with the propagation distance.
      Note, this approach assumes minimal change in the beam's shape during propagation, which is a significant approximation in the variational analysis.
    
	\subsubsection{Linear Chirping}
	
\begin{figure}[h!]
		\centering
		\includegraphics[width=\linewidth]{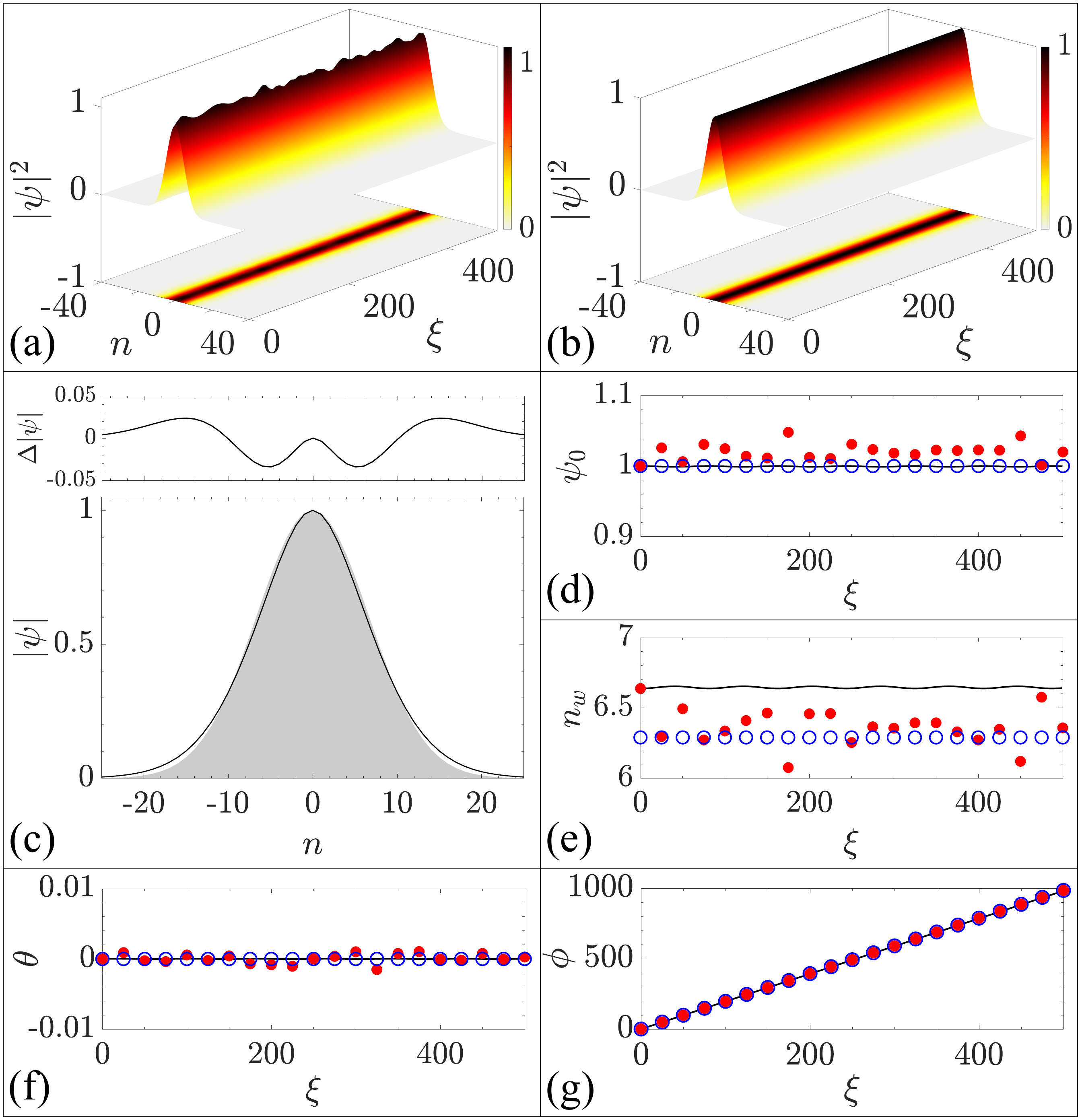}
		\caption{Evolution of (a) a  Gaussian beam as an input of Eq. \eqref{eq:norm_dnlse} and (b) the numerical solution for the DNLSE in a linearly chirped WA under low power conditions ($\mathcal{N}=0$) for $\delta = 10 ~\text{nm}$ and $\alpha_l = 3 \times 10^{-3}$ which makes $n_{sol}= 6.65$.  In plot (c) we superimpose the Gaussian beam shape with the numerical stationary solution of Eq. \eqref{eq:fsolve}. 
		In upper inset we demonstrate how much these two shapes differ.    
		Plot (d)-(g) depict the evolution of beam parameters where the solid line represents the variation predictions. The circle and solid dots represent the corresponding values obtained form the numerical stationary solution and Gaussian feed in respectively.
		}		
		\label{fig:lin_va_num}
	\end{figure}

	In case of  linear chirping, we construct the Lagrangian density corresponding to Eq. \eqref{eq:lin_lin} as,  
	\begin{align}
		\mathcal{L} = &\frac{i}{2} \left( \psi \partial_{\xi} \psi^* - \psi^* \partial_{\xi} \psi \right) + \abs{\partial_n \psi}^2 - 2 \abs{\psi}^2 \nonumber \\
		&+ \alpha_l \abs{n} \left(  2 \abs{\psi}^2 - \abs{\partial_n \psi}^2  \right) . 
	\end{align}
	This Lagrangian density is reduced by inserting the Gaussian \textit{ansatz} given in Eq. \eqref{eq:gauss_ansatz} as, 
	
	\begin{align}
	 	L = \sqrt{\pi} \psi_0^2 \left[ \frac{\alpha_l}{\sqrt{\pi}} \left( -4 n_w^4 \theta^2 + 2 n_w^2 - 1\right) \right. \nonumber \\
		\left. + \frac{n_w^3 }{2}  \left(\theta' + 4 \theta^2 \right)  + n_w \left(\phi' - 2\right) + \frac{1}{2n_w}
		\right],
	\end{align}
	where the prime represents derivative with respect to $\xi$. Employing the  \textit{Euler-Lagrange} equations for the parameters $\nu = \psi_0, n_w, \theta$, and $\phi$, we obtain the set of coupled ODEs expressing the variation of the individual parameter during propagation.  
	\begin{subequations}
	\label{eq:va_lin}
		\begin{align}
			d_{\xi} \psi_0 &= \left( \frac{2\alpha_l}{\sqrt{\pi}} n_w - 1  \right) 2 \psi_0 \theta  			\label{eq:va_lin_psi0}\\
			d_{\xi} n_w	&= - \left( \frac{\alpha_l}{\sqrt{\pi}}n_w - 1 \right) 4 n_w \theta \label{eq:va_lin_nw}\\
			d_{\xi} \theta &= \frac{1}{n_w^4} - 4 \theta^2 + \frac{\alpha_l}{\sqrt{\pi}}\left( 12 n_w \theta^2 - \frac{2}{n_w} - \frac{1}{n_w^3}\right) \label{eq:va_lin_theta}\\
			\label{eq:va_lin_phi}
			d_{\xi} \phi &= 2 - \frac{1}{n_w^2} + \frac{\alpha_l}{\sqrt{\pi}} \left( \frac{3}{2 n_w} - n_w - 2 n_w^3 \theta^2   \right)
		\end{align}
	\end{subequations}
	A steady state solution can be obtained by making $\theta = 0$ in Eq. \eqref{eq:va_lin_psi0} and Eq. \eqref{eq:va_lin_nw} which  leads to a cubic equation for the beam-width as, $n_w^3 + \frac{1}{2}n_w - \frac{\sqrt{\pi}}{2 \alpha_l} = 0$,
	having one real root, 
	\begin{align}
		n_{sol}= -\frac{1}{\tau^{1/3}}+\frac{\tau^{1/3}}{6}, 		
		\label{eq:lin_sol}
	\end{align}
	where $\tau = 6\left(h+\sqrt{6+h^2}\right)$ with $h=9\sqrt{\pi}/\alpha_l$.
	The pair of imaginary roots are ignored as beam width has to be real and positive.
	 A Gaussian beam with specific width $n_{sol}$ and unit amplitude of $\psi_0 = 1$ is seeded as an input for the discrete DNLSE with $\mathcal{N}=0$ (see Eq. \eqref{eq:norm_dnlse}) and its evolution is  illustrated in \figref{fig:lin_va_num} (a). The beam remains nearly stationary with only slight amplitude variations. These deviations occur because the Gaussian function is not a precise solution for the linearly chirped WA, leading to changes in shape during propagation that contradict the fundamental VA assumption. Nonetheless, a stationary solution can be numerically derived through a proposed solution of the form $a_n(\xi) = u(n) e^{i \beta \xi}$, with $\beta$ representing the propagation constant and $u(n)$ as the real mode amplitude distribution complying with the coupled mode equation,
     \begin{align}
		-\beta u(n) + \eta_n^{n+1}u(n+1) + \eta_{n-1}^n u(n-1)=0.
		\label{eq:fsolve}
	\end{align}
To obtain the stationary solution \( u(n) \), the equation can be solved numerically for the appropriate \( \eta_n^{n+1} \) and subsequently used as input in the linear propagation equation with \( \mathcal{N}=0 \). This stationary propagation of the beam is illustrated in Figure \ref{fig:lin_va_num}(b). A comparison between the proposed Gaussian \textit{ansatz} and the numerically obtained stationary solution of Eq. \eqref{eq:fsolve} is shown in Figure \ref{fig:lin_va_num}(c), highlighting the degree of mismatch (\( \Delta |\psi| \)) in the inset. The marginal differences between the two solutions lead to discrepancies evident in Figures \ref{fig:lin_va_num}(d)-(g). These are derived from the governing DNLSE (solid red dots), alongside the predictions of the VA (solid lines) and the full numerical solution (blue circles). The VA provides an effective approximation for predicting the stationary solution of a linearly chirped wave.

	\subsubsection{Quadratic Chirping}
\begin{figure}[h!]
		\centering
		\includegraphics[width=\linewidth]{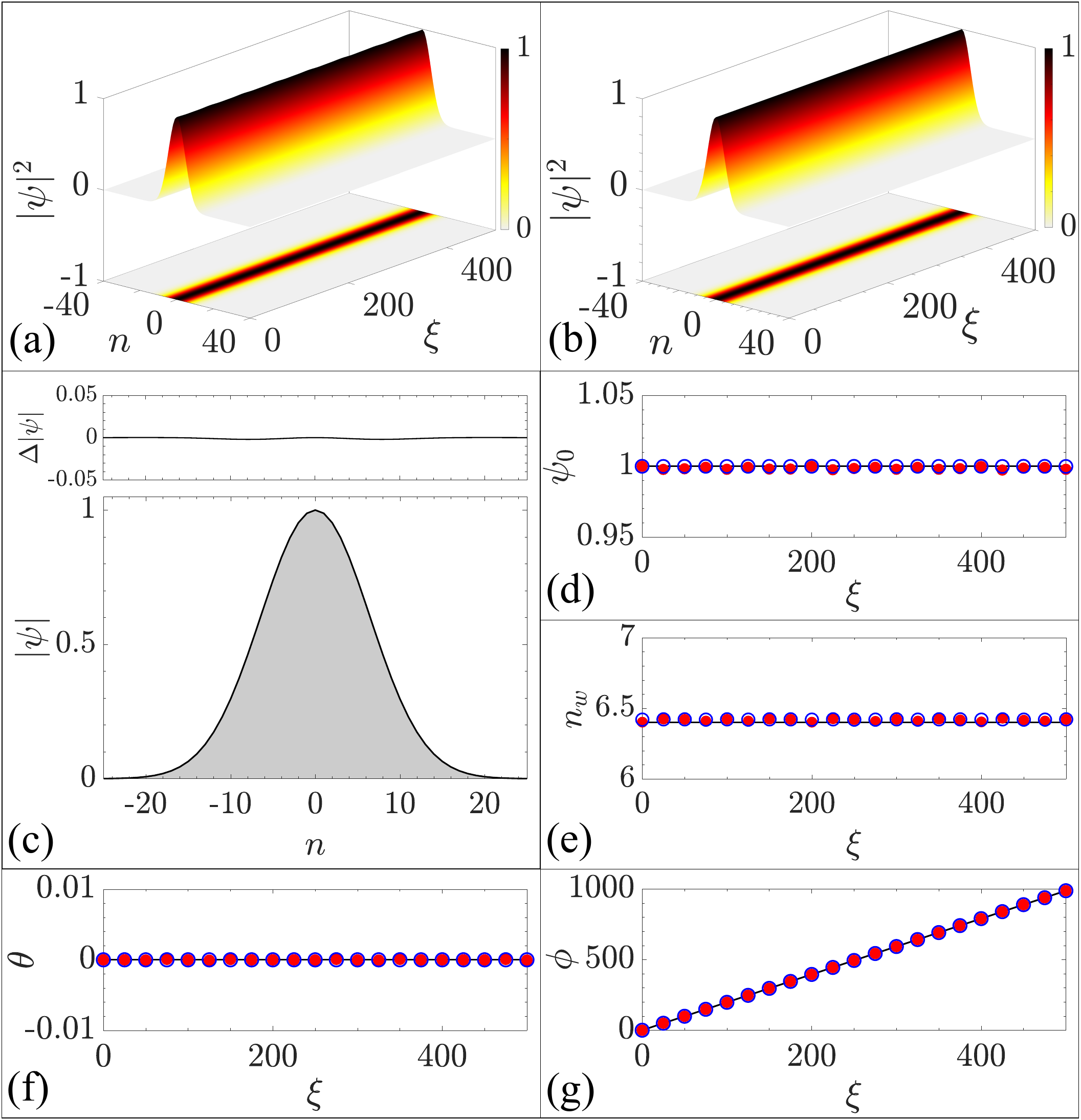}
		\caption{Evolution of (a) A Gaussian beam as an input of Eq. \eqref{eq:norm_dnlse} and (b) the numerical solution for the DNLSEs in a quadratic chirped WA under low power conditions ($\mathcal{N}=0$) for $\delta = 1 ~\text{nm}$ and $\alpha_q = 3\times 10^{-4}$ which makes $n_{sol}=6.4$.  In plot (c) we superimpose the Gaussian beam shape with numerical stationary solution of Eq. \eqref{eq:fsolve}. In upper inset we demonstrate how much these two shapes differ.    
		Plot (d)-(g) depict the evolution of beam parameters where the solid line represents the variation predictions. The circle and solid dots represent the corresponding value obtain from the numerical stationary solution and Gaussian feeding, respectively.
	}		
		\label{fig:quad_va_num}
	\end{figure}

	We extend our investigation to quadratic chirped WA, for which the Lagrangian density can be constructed for Eq. \eqref{eq:lin_quad} as, 
	\begin{align}
		\mathcal{L} = &\frac{i}{2} \left(\psi \partial_{\xi} \psi^* - \psi^* \partial_{\xi} \psi\right) + \abs{\partial_n \psi}^2 - 2\abs{\psi}^2 \nonumber \\
		&+  \alpha_q n^2 \left(2\abs{\psi}^2 - \abs{\partial_n \psi}^2\right). \label{eq:ld_qc}
	\end{align}
	For a Gaussian \textit{ansatz} (see Eq. \eqref{eq:gauss_ansatz}), the reduced Lagrangian is derived as,
	\begin{align}
		L = \sqrt{\pi} \psi_0^2 \left[ \alpha_q \left( -3  n_w^5 \theta^2 +  n_w^3 - \frac{3}{4}  n_w\right)  \right.  \nonumber \\ 
		\left.  +\frac{1}{2}  n_w^3 \left( \theta'+ 4 \theta^2\right) 
		+ n_w\left( \phi'-2\right)+ \frac{1}{2 n_w}\right].
	\end{align}
	Applying the \textit{Ritz's optimization}, we obtain the evolution of beam parameters $\nu = \psi_0,n_w,\theta,$ and $\phi$ in the following coupled ODE form,
	\begin{subequations}
		\begin{align}
			d_{\xi} \psi_0 &=  \left( 3 \alpha_q n_w^2 - 2 \right) \psi_0 \theta \\
			d_{\xi} n_w &= - \left(3 \alpha_q n_w^2 - 2 \right) 2 n_w \theta\\
			d_{\xi} \theta &= \left(3 \alpha_q n_w^2 -1 \right) 4\theta^2 + \frac{1}{n_w^4} - 2 \alpha_q  \label{eq:theta_quad}\\
			d_{\xi} \phi &= 2 - \frac{1}{n_w^2} + \frac{3 \alpha_q}{4} - 3\alpha_q n_w^4 \theta^2
		\end{align}
		\label{eq:quad_va}
	\end{subequations}
	For the quadratic chirped WA we can also find a stationary solution by making $ d_{\xi}\theta = 0$ with the boundary condition $ \theta(0)=0 $ in Eq. \eqref{eq:theta_quad} which leads to the beam width value,  
	\begin{align}
	 n_{sol}= \left( \frac{1}{2 \alpha_q} \right)^{1/4}. \label{eq:quad_va_sol}
	\end{align}	
The evolution of a Gaussian beam with unit magnitude, characterized by the width derived in Eq. \eqref{eq:quad_va_sol}, is presented in \figref{fig:quad_va_num} (a). The findings indicate that the beam propagates steadily, aligning closely with the full numerical solution shown in \figref{fig:quad_va_num} (b). This Gaussian beam serves as a suitable \textit{ansatz} for a waveguide with symmetric quadratic chirping, as evident by \figref{fig:quad_va_num} (c), where it nearly overlays the numerically derived actual solution obtained from Eq. \eqref{eq:fsolve}. The negligible degree of mismatch ($ \Delta |\psi| $), illustrated in the inset of plot (c), underscores the efficacy of the VA, particularly for quadratic chirped waveguides. Additionally, \figref{fig:quad_va_num}(d)-(g) exhibit the evolution of beam parameters under stationary conditions derived from the governing DNLSE (solid red dots), along with predictions from the VA (solid lines) and the full numerical solution (blue circles).

 \begin{figure}[h!]
	\centering
	\includegraphics[width=\linewidth]{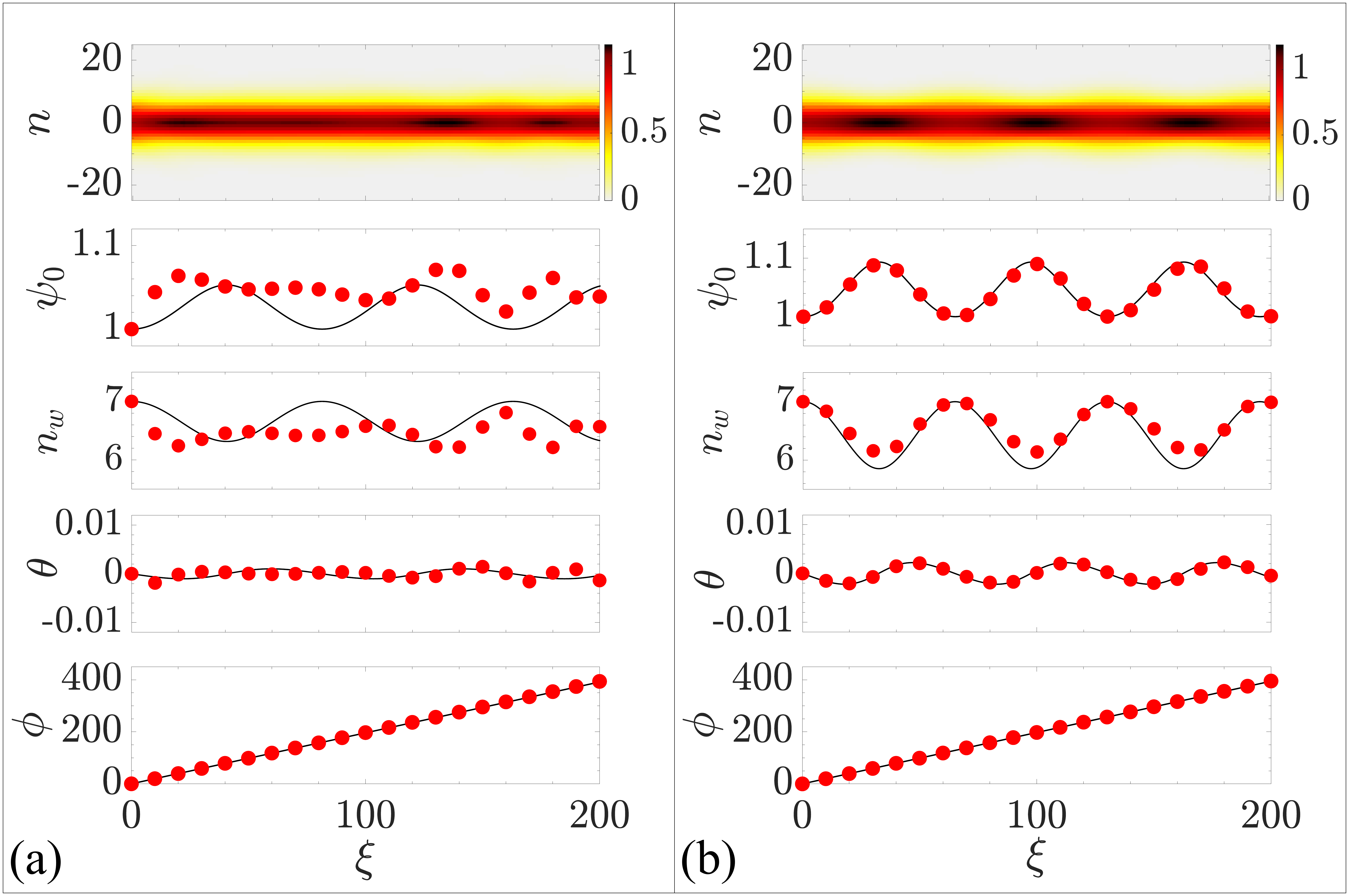}
	\caption{Evolution of a Gaussian beam as an input of DNLSE Eq. \eqref{eq:norm_dnlse} with arbitrary width $n_w = 7 (\neq n_{sol})$ in (a) linear (with $\delta=10$ nm and $\alpha_l=3 \times 10^{-3}$), and (b) quadratic chirped (with $\delta=1$ nm, and $\alpha_q=3 \times 10^{-4}$) WA along with the evolution of the parameters.  Solid red dots denotes the numerical data where predictions from the variational analysis are represented by solid black lines.}
	\label{fig:gen_prop}
\end{figure}

The  VA, which is employed to obtain the stationary solutions in two types of chirped WA can be further extended to investigate the beam dynamics with arbitrary initial conditions. Specifically, the evolution of Gaussian beams with a unit magnitude and an initial width of \( n_w(0) = 7 \) is depicted \figref{fig:gen_prop} (a) and (b) under linear and quadratic chirping, respectively. As illustrated, the comparative analysis of numerical and analytical results shows that VA achieves greater accuracy for quadratic chirp WAs than for linear chirps. This increased precision is attributed to the Gaussian \textit{ansatz} being a better fit for parabolic graded systems \cite{Marcuse:78}, which preserves the beam shape during propagation.  In quadratic waveguide chirping, the beam exhibits periodic oscillations similar to the self-imaging behavior of Gaussian beams in continuous parabolic graded index fibers \cite{karlsson_dynamics_1992}. By prioritizing the effect of the parabolic potential, we can simplify the equations governing the system, leading to the evolution of beam width and phase curvature described by the equations \(d_{\xi} n_w = 4 n_w \theta\) and \(d_{\xi} \theta = -2\alpha - 4\theta^2 + 1/n_w^4\). This simplification permits the representation of the evolution of beam width \(n_w\) as a second-order differential equation as,
    \begin{align}
		d^2_{\xi} n_w + 8 \alpha n_w - \frac{4}{n_w^3} = 0.
	\end{align}
	This expression is  analogous to the evolution of a Gaussian beam width ($w$) in a continuous graded index medium which is given as $d^2 w/dz^2 + b^2 w - 1/k^2 w^3 = 0$, where $b$ is the index gradient and $k$ is the wavenumber constant \cite{Agrawal_Physics_2023}.
	Drawing an analogy between these two expressions by making $b = 2\sqrt{2\alpha_q}$ and $k = 1/2$, the evolution of the beam width inside the quadratically chirped WA (with $\alpha_q=3 \times 10^{-4}$) turns out to be $n_w(\xi) = n_w(0)\sqrt{f(\xi)}$, where $f(\xi) = \cos^2(2\pi\xi/L_p) + C_f^2 \sin^2(2\pi\xi/L_p)$, with $L_p = \pi/\sqrt{2 \alpha_q}$, $C_f = (n_{\text{sol}}/n_w(0))^2$ and $n_{\text{sol}} = 1/\sqrt{kb} = (1/2\alpha_q)^{1/4}$.
	This expression allows us to determine the self-imaging period where the beam inside the WA is restored to its original width after propagating for a distance of $\xi = L_p/2 \approx 65$, which corroborates well with the numerical results shown in \figref{fig:gen_prop} (b).
	Inside a WA, a discrete Gaussian beam  of arbitrary  width can therefore be focused  at a distance of $L_p/4$ and this length can be manipulated by the chirping strength.

	\subsection{Nonlinear Case (in presence of Kerr nonlinearity)}
	
    In high beam power scenarios within a WA, nonlinear phenomena such as the Kerr effect cannot be disregarded. This nonlinear effect produces SPM, which facilitates the self-focusing of the beam. This process counteracts beam spreading caused by diffraction, ultimately creating conditions conducive to the emergence of a discrete soliton \cite{christodoulides_discrete_1988,morandotti_self-focusing_2001}. In a uniform WA, the dynamics of the beam is governed by the normalized DNLSE, where equilibrium is achieved with a normalized parameter of $\mathcal{N} = 1$, corresponding to a typical sech-type solution shown in Eq. \eqref{eq:sechsol}
    
    \subsubsection{Evolution dynamics of a Gaussian beam in a nonlinear quadratic chirped WA}

\begin{figure}[h!]
	\centering
	\includegraphics[width=\linewidth]{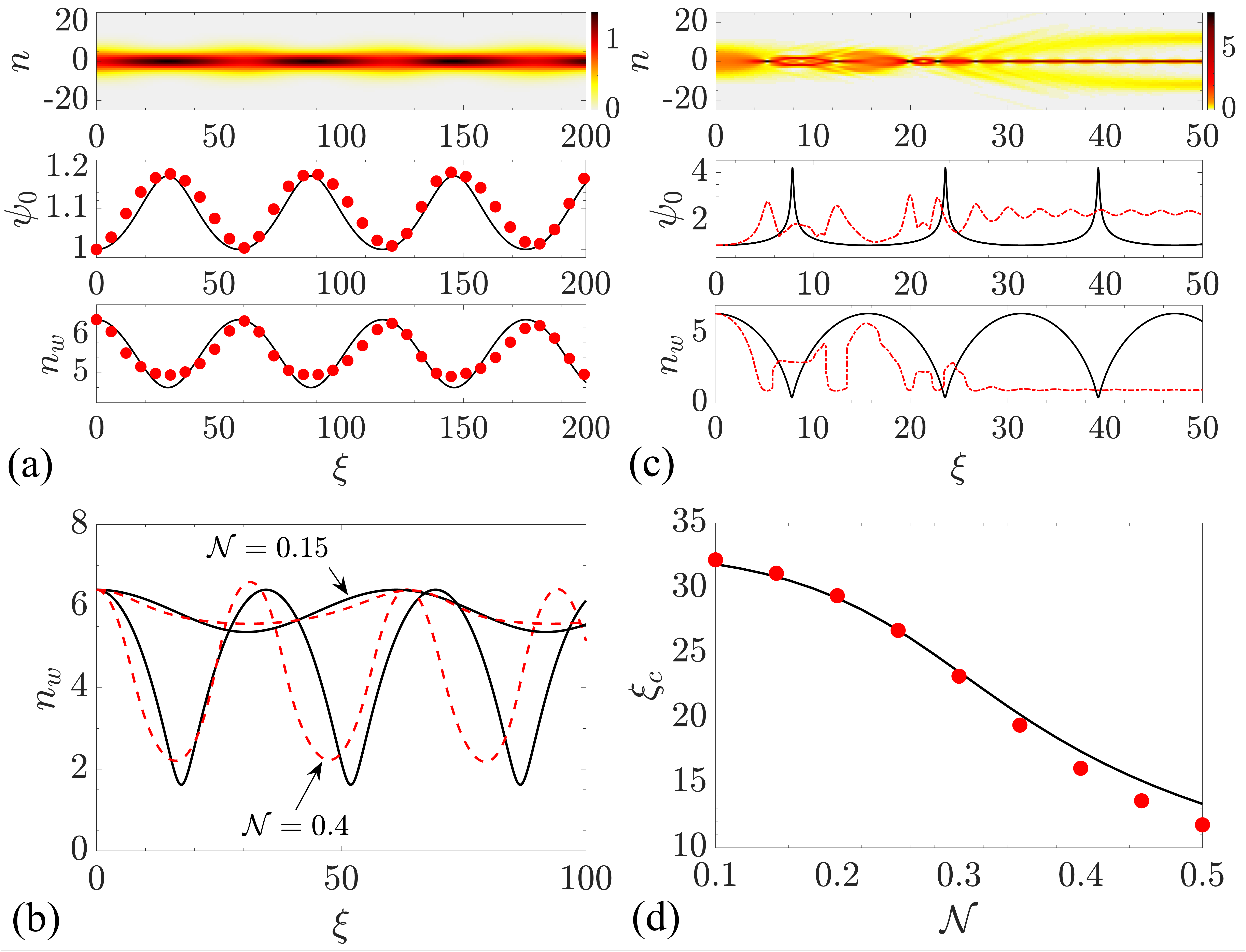}
	\caption{Evolution of a Gaussian beam as a seeded input of Eq. \eqref{eq:norm_dnlse} under the influence of non-vanishing Kerr effect with (a) $\mathcal{N} = 0.2$ and (c) $\mathcal{N} = 0.8$. In the inset we demonstrate how amplitude ($\psi_0$) and beam-width ($n_w$) are affected by SPM. For relatively low power level ($\mathcal{N} = 0.2$) at which the distortion of the Gaussian beam shape is tolerable,  variational analysis (solid lines) agrees well with numerical data (solid dots). In plot (b) we portray the variation of oscillatory beam-width for two different $\mathcal{N}$ values where variational analysis (solid lines) are corresponding numerical data (dotted lines) are compared. In plot (d) we show how the oscillation period ($\xi_c$) decreases with increasing nonlinearity ($\mathcal{N}$) and the variational prediction (solid line) is consistent with full numerical analysis (solid dots).   }
	\label{fig:gauss_quad_nonl}
\end{figure}

\begin{figure}[ht!]
		\centering
		\includegraphics[width=\linewidth]{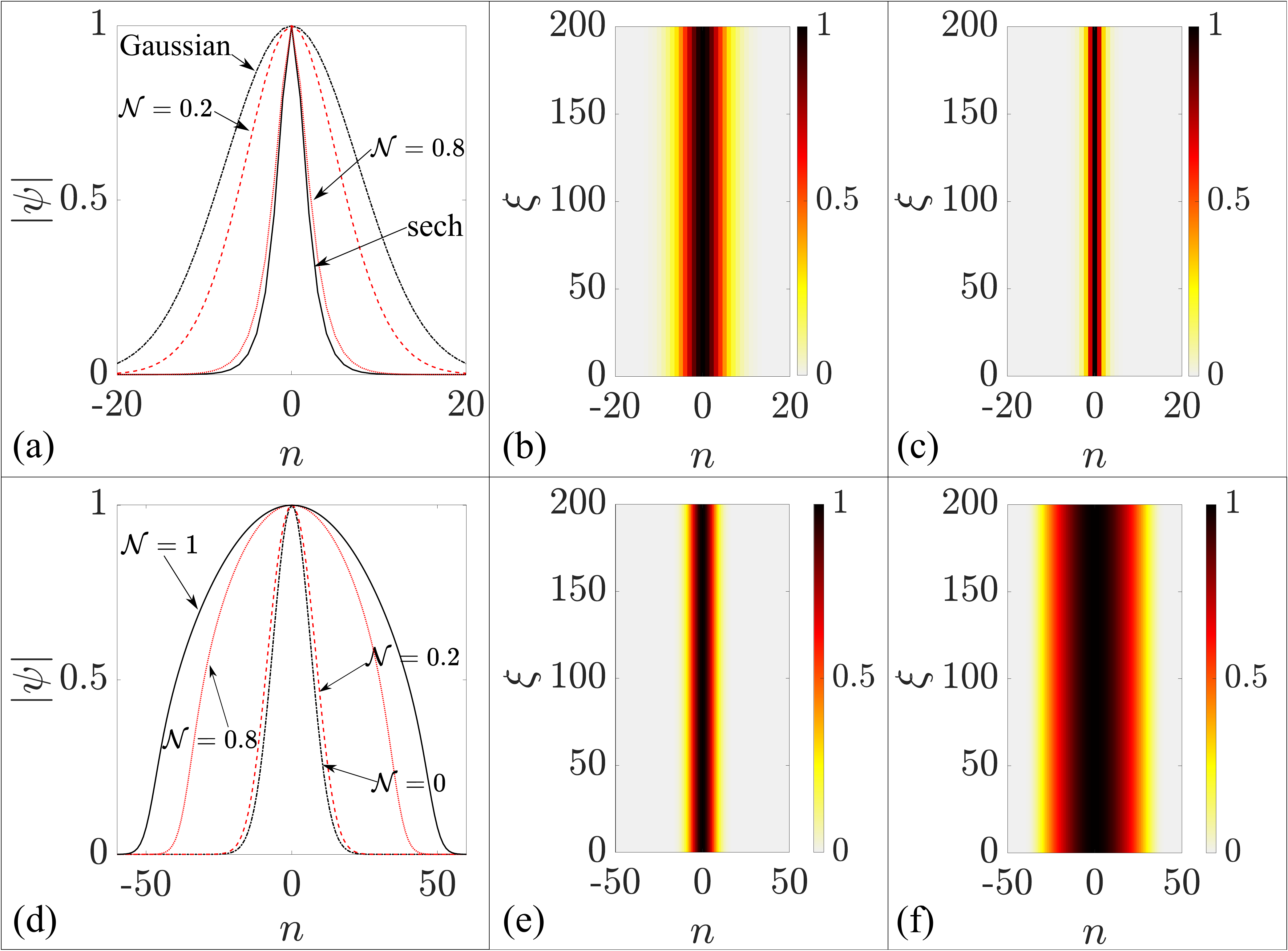}
		\caption{ (a) We compare the full numerical solution with  Gaussian solution (dashed dotted) as per Eq. \eqref{eq:quad_va_sol} for low nonlinearity ($\mathcal{N} = 0.2$) and for high nonlinearity ($\mathcal{N} = 0.8$) with sech-type shape which is a natural solution in a uniform WA.  Evolution of the stationary solution for (b) $\mathcal{N} = 0.2$  (dashed line in (a)) and (c) $\mathcal{N} = 0.8$ (dotted line in (a)) under a chirping strength of $\delta = 1 \text{nm}$. {(d) Comparision of stationary beam shape under defocusing nonlinearity for different $\mathcal{N} $ values. Propagation plot for (e) $\mathcal{N}=0.2$ and (f) $\mathcal{N}=1. $}}
		\label{fig:nonl_sols}
	\end{figure}

    This section examines the influence of nonlinearity on a Gaussian beam, which serves as a suitable \textit{ansatz} function in a linear WA. The governing equation for beam propagation within a nonlinear quadratic chirped WA is presented in Eq. \eqref{eq:cont_quad_2}. Utilizing the VA, the Gaussian \textit{ansatz} outlined in Eq.\eqref{eq:gauss_ansatz} leads to a set of ODE,
    \begin{subequations}
		\begin{align}
			d_{\xi} \psi_0 &=  \left( 3 \alpha_q n_w^2 - 2 \right) \psi_0 \theta \\
			d_{\xi} n_w &= - \left(3 \alpha_q n_w^2 - 2 \right) 2 n_w \theta\\
			d_{\xi} \theta &= \left(3 \alpha_q n_w^2 -1 \right) 4\theta^2 + \frac{1}{n_w^4} - 2 \alpha  - \frac{\mathcal{N}^2}{2\sqrt{2}}\frac{\psi_0^2}{n_w^2} \label{eq:nonl_theta_quad}\\
			d_{\xi} \phi &= 2 - \frac{1}{n_w^2} + \frac{3 \alpha_q}{4} - 3\alpha_q n_w^4 \theta^2 + \frac{5 \mathcal{N}^2}{4\sqrt{2}}\psi_0^2
		\end{align}
		\label{eq:nonl_quad_va}
	\end{subequations}
Exploiting the set in Eq.\eqref{eq:nonl_quad_va} we intend to examine the dynamics of a Gaussian beam inside quadratic chirped WA where Kerr nonlinearity should play a crucial role. In  \figref{fig:gauss_quad_nonl}(a) we illustrate the beam oscillation owing to the SPM induced  self-focusing for a moderate power level of $P_0=270$ W (which makes  $\mathcal{N} = 0.2$). The variational results (solid red dots) are found to be in well agreement with full numerical results (solid black line). 
	Note that this oscillatory beam dynamics bears a resemblance to the self-focusing observed in parabolic graded index fibers \cite{karlsson_dynamics_1992, Hansson2020}. The evolution of the beam-width is approximately governed by the following ordinary second order differential equation, 
	\begin{align}
		d_\xi^2 n_w + 8 \alpha_q n_w - \frac{4}{n_w^3} - \sqrt{\frac{2}{\pi}}\mathcal{N}^2 E \frac{1}{n_w^2} = 0
		\label{eq:nw}
	\end{align}
	where $E = \int_{-\infty}^{\infty}\abs{\psi}^2 dn = \sqrt{\pi}\psi_0(0)^2 n_w(0)$ is the total energy which is conserved.  The ODE presented in Eq. \eqref{eq:nw} effectively accounts for the oscillatory behavior of a Gaussian beam under moderate nonlinearity (with $\mathcal{N} \leq 0.5$), as evidenced by the good correspondence between analytical results (solid black lines) and full numerical simulations (red dashed line) shown in \figref{fig:gauss_quad_nonl}(b). An increase in power leads to significant beam compression, eventually distorting the Gaussian shape as it encompasses only a few waveguides. This strong confinement results in the emergence of a Peierls-Nabarro (PN) potential barrier in nonlinear lattices, which subsequently forms a stop-band for localized modes in strongly nonlinear waveguides \cite{kivshar_peierls-nabarro_1993,papacharalampous_soliton_2003,lara_collision_2023}. A high nonlinearity ($\mathcal{N} = 0.8$) further constrains the beam to the central waveguide, causing notable distortion of the beam shape and discrepancies between variational results and numerical data, as shown in \figref{fig:gauss_quad_nonl} (c). The analysis of oscillatory dynamics across various nonlinearity levels reveals a decreased oscillation period ($\xi_c$) with increasing $\mathcal{N}$, which is corroborated by the agreement between analytical predictions and numerical results in \figref{fig:gauss_quad_nonl} (d). While a stationary Gaussian solution exists for $\mathcal{N}=0$, the same is not true for non-zero nonlinearity, leading to beam collapse at high nonlinear levels. Efforts are ongoing to identify a stationary solution in the form $a_n(\xi) = u(n)e^{-i \beta \xi}$ through its incorporation into the DNSLE,

     \begin{align}
		-\beta u(n) + \eta_n^{n+1}u(n+1) + \eta_{n-1}^n u(n-1) \nonumber \\ + \mathcal{N}^2 u(n)^3=0.
		\label{eq:fsolve_nl}
	\end{align}
	In this study, we numerically solve the equation for a quadratic chirped WA across various nonlinearity values ($\mathcal{N}$). The results, depicted in \figref{fig:nonl_sols} (a), illustrate that as nonlinearity increases, the steady-state solution's shape becomes increasingly compressed in the transverse coordinate ($n$-space) and aligns with the sech-type function, which represents a soliton solution as shown in the propagation plot \figref{fig:nonl_sols} (b) and (c). In the limit as $\mathcal{N}$ approaches 1, the sech-type beam demonstrates greater stability compared to a Gaussian beam. Consequently, we adopt the sech-type beam as our \textit{ansatz} for investigation in the dynamics of the chirped WA.  {We further analyze the formation of solitonic beam under defocusing nonlinearity with in positively chirped WA and the results are depicted in \figref{fig:nonl_sols} (d)-(f). Notably for high $\mathcal{N}$ values the beam shape differs significantly from Gaussian. }

	\subsection{Linear stability analysis of the stationary solutions}
\begin{figure}[h!]
	\centering
	\includegraphics[width=\linewidth]{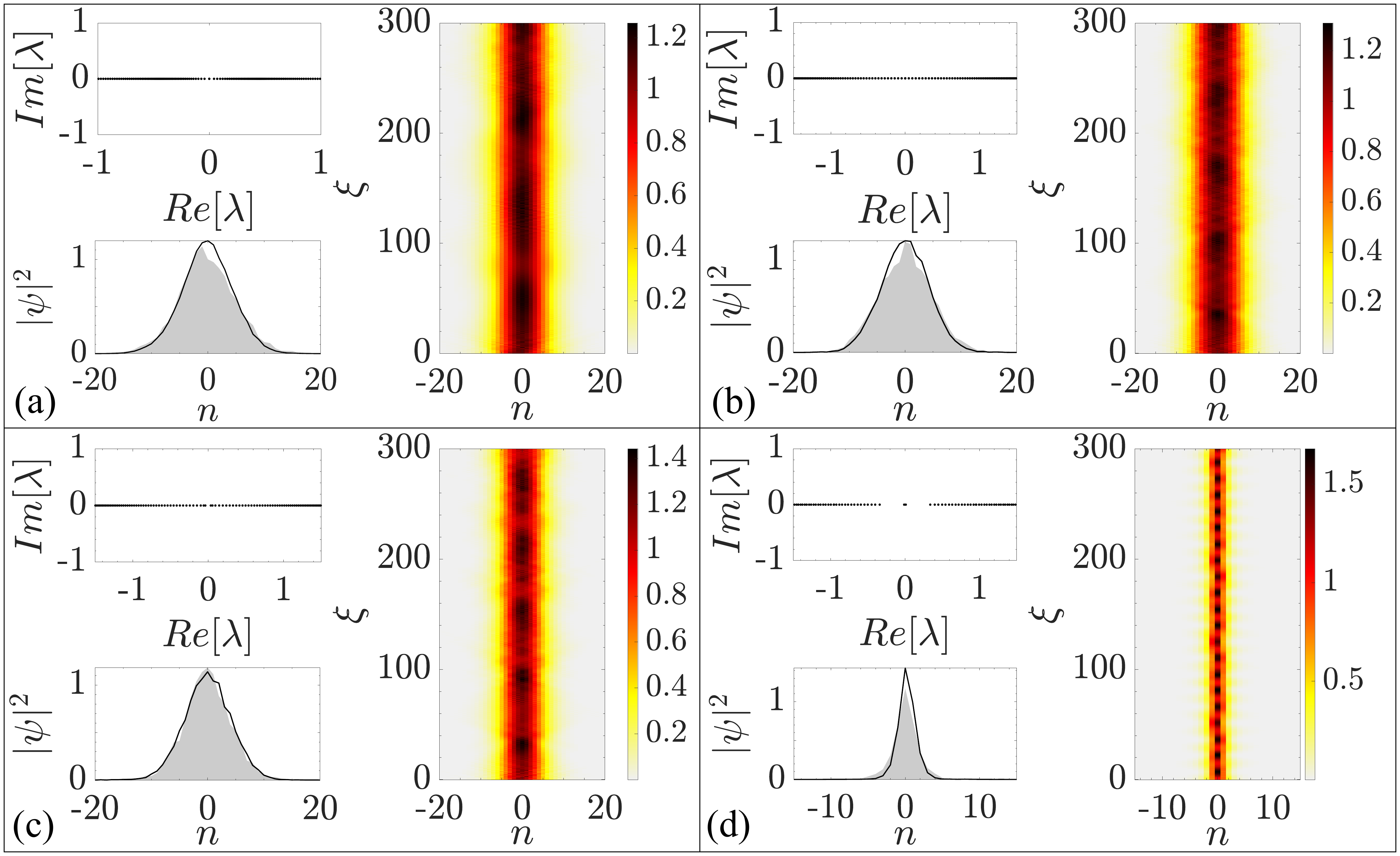}
	\caption{Evolution of numerical solution in the absence of Kerr self-focusing, obtained under (a) linear and (b) quadratic chirping for $\mathcal{N} = 0$, and for (c) $\mathcal{N} = 0.2$ and (d) $\mathcal{N} = 0.8$ under quadratic chirping with a random noise with an amplitude of 10\% of the peak solution amplitude. The growth parameter $Im(\lambda)$ is found to be zero for all the cases confirming the stability of the solution. The insets also demonstrate the superposition of the input and output beam showing marginal distortion under input noise. }
	\label{fig:stab_check}
\end{figure}

	In this section, we perform a linear stability analysis (LSA) for all the stationary solutions that we obtained in previous sections. This analysis is conducted by introducing a small amplitude perturbation to the stationary solution as, 
		\begin{align}
			a_n(\xi) = \left[u_n + v_n e^{-i \lambda \xi} + w_n^* e^{i \lambda^* \xi}\right] e^{i \beta \xi},
		\end{align}
		where, $v_n$ and $w_n$ are small amplitude complex perturbations to the stationary solution $u_n$, with complex propagation constants $\lambda$ \cite{sukhorukov_generation_2003}.
		The substitution of this field into the governing DNLSE (see Eq. \eqref{eq:norm_dnlse}), and respective linearization of  $v_n$ and $w_n$ (as $v_n,w_n \ll u_n$), leads to the eigenvalue problem
		\begin{align}
			\mathcal{M} X=\lambda X
		\end{align}
		where $X = [V; W]$ with $V = [v_{-N}; v_{-N +1 }; ... v_0; ... , v_{(N-1)}; v_N]$, $W = [w_{-N}; w_{-N + 1}; ... w_0; ..., w_{(N-1)}; w_N]$ and $\mathcal{M}$ is a $2(2N + 1) \times 2(2N + 1)$ matrix given as 
		\begin{align}
			\fontsize{7}{5}
			\mathcal{M} \hspace*{-3pt} = \hspace*{-3pt}\begin{bmatrix}
				D_{-N} & -\eta_{-N}^{-N+1} & 0 &. &. & . & . & 0 \\
				. &. & . & . & . & . & . & .\\
				. & -\eta_{n-1}^n & D_n & -\eta_n^{n+1} &. &. &. & .\\
				. &. & . & . & . & . & . & .\\
				. & . & -\eta_{N-1}^N & D_N & 0  & 0 &. & . \\
				. & . & 0 & 0 &  G_{-N} & \eta_{-N}^{-N+1} &. &.\\
				. & . & . & . & . & .  &. & .\\
				. & . & . & . & . & .  &. & .\\
				. & . &. & . &\eta_{n-1}^n & G_n & \eta_n^{n+1} & .\\
				. & . & . & . & . & . & . & .\\
				0 & . & . & . & 0 & 0 &\eta_{N-1}^N & G_{N} 
			\end{bmatrix}
		\end{align}
	with $D_n = \beta - \mathcal{N}^2 \left(2\abs{u_n}^2 + u_n^{*2}\right)$,  $G_n = -\beta 
		+ \mathcal{N}^2 \left(2 \abs{u_n}^2 + u_n^2\right)$, and $(2N + 1)$ is the total number of waveguides. The off-diagonal elements of the matrix $\mathcal{M}$ corresponds to the coupling coefficients and are determined by the corresponding chirping scheme of the system, while the diagonal terms are derived from the propagation constant ($\beta$) and SPM ($\mathcal{N}^2 (2 \abs{u}^2 + u_n^{*2})$) terms of the stationary solution $u_n$.
		The matrix form reduces the determination of the localized solutions of $v_n$ and $w_n$ in to a eigenvalue problem with $\lambda = \text{Eig}(\mathcal{M})$. Eigenvalues with $Im(\lambda) \neq 0 $  indicate the instability growth where the perturbation grows with propagation as $v_n$ or $w_n$ increase exponentially and the stationary solution is unstable.
The findings of LSA, summarized in \figref{fig:stab_check}, indicate that all four cases exhibit purely real eigenvalues, confirming their stability against perturbations \cite{sukhorukov_generation_2003}. This stability was further validated through numerical simulations of the DNLSE Eq.  \eqref{eq:norm_dnlse} with inputs affected by 10\% amplitude noise. As illustrated in \figref{fig:stab_check}, the simulations corroborate with the LSA predictions regarding stability. Notably, two discrete points in the eigenvalue spectrum shown in \figref{fig:stab_check} (d) correspond to oscillating modes, consistent with numerical results \cite{Pelinovsky1998}.

\subsection{Dynamics of discrete soliton in chirped WA}

As observed in the earlier section, a sech-type beam, resembling a discrete soliton under the Kerr effect, emerges as the exact solution. Consequently, we adopt a sech-type shape as our \textit{ansatz} for the VA. It is important to note that the results from VA are highly dependent on the selection of the \textit{ansatz} function, which should maintain its shape during propagation under perturbations. 
 The governing equation for optical field inside a chirped WA system can thus be formulated as, 
 
	\begin{align}
		i \partial_{\xi} \psi + 2 \psi + \partial_n^2 \psi + \abs{\psi}^2 \psi = i \epsilon,
		\label{eq:pert_NLSE}
	\end{align}
	where $\epsilon$ accounts for the perturbation owing to waveguide chirping. The corresponding Lagrangian density $\mathcal{L}$ for Eq. \eqref{eq:pert_NLSE} is constructed as, 
	\begin{align}
		\mathcal{L} = \frac{i}{2}\left( \psi \partial_{\xi} \psi^* - \psi^* \partial_{\xi} \psi \right) - 2 \abs{\psi}^2 + \abs{\partial_n \psi}^2 - \frac{1}{2}\abs{\psi}^4 \nonumber \\
		+  i \left(\epsilon \psi^* - \epsilon^* \psi \right).
	\end{align}
	We consider the \textit{ansatz} as \cite{tran_diffractive_2013},
	\begin{align}
		\psi(n,\xi) = \psi_0 \sech \left[\frac{\psi_0 (n - n_0)}{\mathcal{D}} \right]e^{i\phi + i \kappa_0(n - n_0)},
	\end{align}
	where $\psi_0$, $n_0$, $\kappa_0$, and $\phi$ are the soliton parameters representing amplitude, peak position, central wavenumber, and phase respectively, and $\mathcal{D} = \sqrt{2 \cos(\kappa_0(0))}$. We reduce the Lagrangian by adopting the \textit{ansatz} function as  $L = \int_{-\infty}^{\infty}\mathcal{L} dn $, and get 
    \begin{align}
		L = &\left(d_{\xi}\phi - \kappa d_{\xi}n_0 + \kappa^2 - 2  \right) 2 \mathcal{D} \psi_0 \nonumber \\
		& \frac{2}{3}\left(\frac{1}{\mathcal{D}} - \mathcal{D} \right) + \int_{-\infty}^{\infty} i (\epsilon \psi^* - \epsilon^* \psi)dn
		\label{eq:red_lagr}
	\end{align}
    Employing the \textit{Euler-Lagrangian equation} \cite{anderson_pereira-stenflo_1999} we finally obtain the following set of ODE showing how the beam parameters are affected due to chirping in the arrangement of a WA.
 	\begin{subequations}
 		\label{eq:va_DS}
		\begin{align}
			d_{\xi}\psi_0 &= \frac{1}{2 \mathcal{D}}P_{\phi} \\
			d_{\xi} \kappa_0 &= - \frac{1}{2\mathcal{D} \psi_0} \left(\kappa_0 P_{\phi}+ P_{n_0}\right)  \\
			d_{\xi} n_0 &= 2 \kappa_0 + \frac{1}{2 \mathcal{D} \psi_0}P_{\kappa_0} \\
			d_{\xi} \phi &= 2 + \psi_0^2 + \kappa_0^2 - \frac{\psi_0^2}{\mathcal{D}^2} \nonumber \\ &+ \frac{1}{2 \mathcal{D} \psi_0} \left(\kappa_0P_{\kappa_0} - \psi_0 P_{\psi_0} \right).	
		\end{align}
	\label{eq:VAgen}
	\end{subequations}	
	Here, $P_{\nu}$  are defined as $P_{\nu} = \int_{-\infty}^{\infty}i \left(\epsilon \partial_\nu\psi^* - \epsilon^* \partial_\nu \psi\right)dn$, which can be evaluated separately for each parameter $\nu$ under the appropriate perturbation  $\epsilon$.

	\subsubsection{Linear Chirping}
\begin{figure}[h!]
		\centering
		\includegraphics[width=\linewidth]{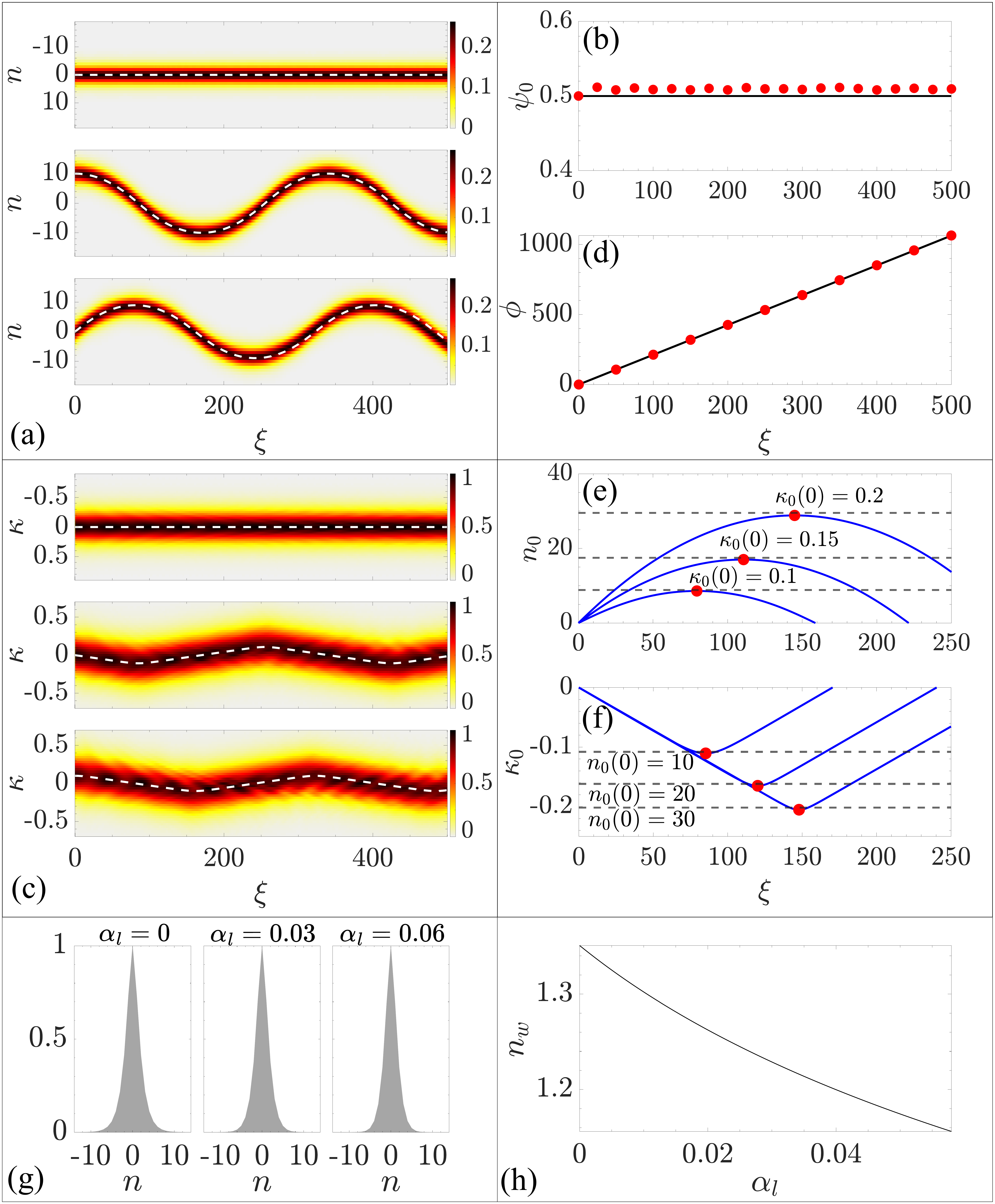}
		\caption{Evolution of a DS under linear chirp for three initial cases (i) $n_0(0) = 0$ and $\kappa_0(0) = 0$, (ii) $n_0(0) =10$ and $\kappa_0(0) = 0$, and (iii) $n_0(0) = 0$ and $\kappa_0(0) = 0.1$ for chirping strength of $\delta = 2.5~\text{nm}$.
			The evolution of the soliton in the (a) $n$-space and (c) $\kappa$-space is are shown for each case with the prediction of the central location from the variational analysis given by dashed white lines.
			The corresponding evolution of the (b) amplitude and (d) phase as per the DNLSE (solid red cirles) and variational analysis (solid lines), respectively. Plot (e) and (f) shows the trace of beam position in $n$ and $\kappa$ space for different $n_0 (0)$ and $\kappa_0(0)$ values. The beam location limit (red dots on the curve) agrees well with analytical prediction (horizontal dotted lines). 
			{(g) Soliton structure for three different chirping strengths. (f) Dependency of soliton width with $\alpha_l$.}
		}
		\label{fig:lin_ds}
	\end{figure}

	In case of linearly chirped WA, where the beam dynamics is governed by Eq. \eqref{eq:cont_linear_2} the perturbation term due to chirping is given by, 
	\begin{align}
		\epsilon_l = -i \alpha_l \left[\abs{n} 2 \psi + \abs{n}\partial_n^2 \psi + \sgn(n) \partial_n \psi\right].
	\end{align}
	For this given $\epsilon_l$, $P_\nu$ are obtained as, 
	\begin{subequations}
		\label{eq:va_lin_ds}
		\begin{align}
			P_{\phi} &= 0
			\\
			P_{n_0} &= -\frac{2\alpha_{l}\mathcal{D}^2}{3} \mathcal{A} \chi_t \left[\mathcal{A}^2 \chi^2_t  + 3\mathcal{B}\right]
			\\
			P_{\kappa_0} &= 4 \alpha_{l}\mathcal{D}^2\kappa_0 \log \left[\frac{\chi_s}{2}\right] 			
			\\
			P_{\psi_{0}}&=-2\alpha\mathcal{D}n_{0}\chi_t\left[\frac{1}{3}\mathcal{A}^{2}\chi_t^2 + \mathcal{B}\right]
			\nonumber \\
			& - \frac{4\alpha_{l} \psi_0}{3}\left[\frac{\chi_s^2}{2} +  \log \left(2/\chi_s\right)  \right]
		\end{align}
	\end{subequations}
	where $\mathcal{A} = \psi_0/\mathcal{D}$, and $\mathcal{B}= (\kappa_0^2 - 2)$, with $\chi_t=\tanh(\mathcal{A}n_0)$ and  $\chi_s=\sech(\mathcal{A}n_0)$.
	Note,  one can obtain the equation of motion for the different beam parameters under linear chirping by plugging  Eq. \eqref{eq:va_lin_ds} in Eq. \eqref{eq:va_DS}.
For a chirping strength of $\delta = 2.5~\text{nm}$, the beam dynamics and evolution of various beam parameters are illustrated in \figref{fig:lin_ds} (a)-(d). The results obtained from the VA are compared with those from the DNLSE under three distinct sets of initial conditions: (i) $n_0 = 0$ and $\kappa_0 = 0$, (ii) $n_0 = 10$ and $\kappa_0 = 0$, and (iii) $n_0 = 0$ and $\kappa_0 = 0.1$. The beam dynamics in $n$ and $\kappa$ space are depicted in \figref{fig:lin_ds} (a) and (c) respectively, showcasing the variational predictions (white dashed lines) from Eq.~\eqref{eq:va_lin_ds} alongside numerically obtained mesh plots from the DNLSE in Eq. \eqref{eq:norm_dnlse}. The oscillatory behavior of the propagating beam aligns closely with variational predictions across various initial conditions, where the beam is launched from different waveguide channels ($n_0$) and transverse wavenumbers ($\kappa_0$). Further, amplitude ($\psi_0$) and phase ($\phi$) variations, represented in \figref{fig:lin_ds} (b) and (d), illustrate strong agreement between the VA and numerical results.  Notably, the influence of derivative terms in $\epsilon_l$ is minimal under low chirping strength, allowing for simplified equations for beam position and wavenumber dynamics, given as
	\begin{subequations}
		\begin{align}
			d_\xi n_0 &= 2 \kappa_0 \\
			d_\xi \kappa_0 &= -2 \alpha_l \tanh\left(\mathcal{A} n_0\right).
		\end{align}
		\label{eq:va_lin_ds_2}
	\end{subequations}
Set of Eq.\eqref{eq:va_lin_ds_2} leads to the compact equation, $d_\xi^2 n_0 = -4\alpha_l \tanh\left(\mathcal{A} n_0\right)$.
The closed-form solution for the differential equation remains elusive. However, Eq. \eqref{eq:va_lin_ds_2} allow insight into beam dynamics. For a beam launched normally from the central waveguide, the transverse wavenumber evolves to a maximum value given by $\kappa_{max} = {\sqrt{\frac{2 \alpha_l}{\mathcal{A}} \log\left(\cosh \left[ \mathcal{A} n_0(0) \right] \right)}}$. Conversely, when launched at an angle from the central waveguide, the beam is confined within a limiting channel, defined by $n_{max} = {\frac{1}{\mathcal{A}}\cosh^{-1}\left[\exp\left(\frac{\mathcal{A} \kappa_0^2(0)}{2 \alpha_l}\right)\right]}$. In the limit where both $n_0(0)$ and $\kappa_0(0)$ approach zero, the beam propagates straight without oscillations, aligning with numerical simulations. \figref{fig:lin_ds} (e),(f) demonstrate the beam's position during propagation in $n$ and $\kappa$ space for various launch angles and sites, with numerical results corroborating analytical predictions. {In \figref{fig:lin_ds} (g) and (h) we finally demonstrate the marginal influence of chirping strength $\alpha_l$ on soliton beam-width.}

	\subsubsection{Quadratic Chirping}
\begin{figure}[h!]
		\centering
		\includegraphics[width=\linewidth]{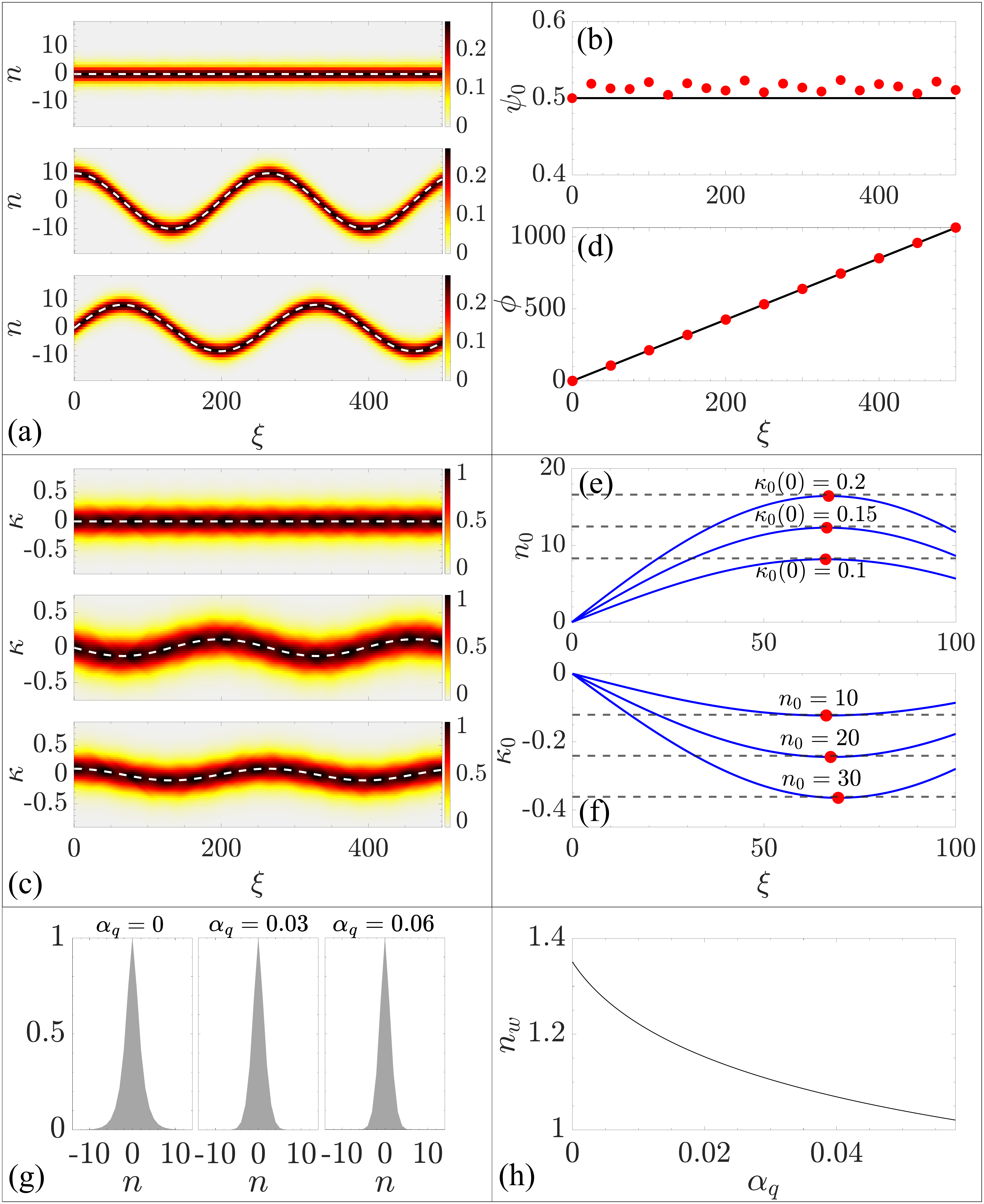}
		\caption{Evolution of a DS under quadratic chirp for three initial cases (i) $n_0(0) = 0$ and $\kappa_0(0) = 0$, (ii) $n_0(0) =10$ and $\kappa_0(0) = 0$, and (iii) $n_0(0) = 0$ and $\kappa_0(0) = 0.1$ for chirping strength of $\delta = 0.25~\text{nm}$ ($\alpha_q=7.25 \times 10^{-5}$).
	The evolution of the soliton in the (a) $n$-space and (c) $\kappa$-space is are shown for each case with the prediction of the central location from the variational analysis is given by dashed white lines.
	The corresponding evolution of the (b) amplitude and (d) phase as per the DNLSE (solid red cirles) and variational analysis (solid lines), respectively.  Plot (e) and (f) shows the trace of beam position in $n$ and $\kappa$ space for different $n_0 (0)$ and $\kappa_0(0)$ values. The beam location limit (red dots on the curve) agrees well with analytical prediction (horizontal dotted lines).  {(g) Soliton structure for three different chirping strengths. (h) Dependency of soliton width with $\alpha_q$.} }
		\label{fig:quad_ds}
	\end{figure}

	Next, we consider a quadratic chirped WA where the perturbation due to chirping is introduced in the governing Eq.~\eqref{eq:cont_quad_2} as,  
\begin{align}
	\epsilon_q = - i \alpha_q \left( 2n^2 \psi -  n^2 \partial_n^2 \psi -  2  n \partial_n \psi\right).
\end{align} 
Implementing  $\epsilon_q$ in the NLSE, Eq. \eqref{eq:pert_NLSE}  we derive $P_{\nu}$, 
	\begin{subequations}
		\label{eq:va_quad_ds}
		\begin{align}
			P_{\phi} &= 0 \\
			P_{\kappa_0} &= - 4\alpha_q \mathcal{D} \psi_0 \left( n_0^2 + \frac{\pi^2}{12}\frac{1}{\mathcal{A}^2} \right) \\
			P_{n_0} &= -4 \alpha_q \mathcal{D} \psi_0 n_0 \left(\frac{\mathcal{A}^2}{3} + \mathcal{B}\right) \\
			P_{\psi_0} &= 2\alpha_q \mathcal{D} \left[ \frac{\pi^2}{12}\left( \frac{\mathcal{B}}{\mathcal{A}^2} -\frac{1}{3} \right) + \left(\mathcal{A}^2 - \mathcal{B}\right) n_0^2 - \frac{1}{3}  \right].
		\end{align}
	\end{subequations}	
     The equation of motion for various beam parameters is derived by substituting Eq. \eqref{eq:va_quad_ds} into Eq. \eqref{eq:va_DS}. At a chirping strength of $\delta = 0.25 ~\text{nm}$, a comparison of variational results and the DNLSE is presented in \figref{fig:quad_ds} for three initial conditions: (i) $n_0 = 0$ and $\kappa_0 = 0$, (ii) $n_0 = 10$ and $\kappa_0 = 0$, and (iii) $n_0 = 0$ and $\kappa_0 = 0.1$.
	The variational results (white dashed lines) accurately trace the beam location in $n$ (plot \figref{fig:quad_ds} (b)) and $\kappa$ (\figref{fig:quad_ds} (c)) space, respectively. Additionally, the evolution of $\psi_0$ and $\phi$ is illustrated in \figref{fig:quad_ds} (b) and (d). The set of Eq. \eqref{eq:VAgen} is simplified by omitting the last two terms in $\epsilon_q$, which are deemed negligible, resulting in a more concise set of ODE,
	\begin{subequations}
		\begin{align}
			d_{\xi} n_0 &= 2\kappa_0 \\
			d_{\xi} \kappa_0 &= -4\alpha_q n_0.
		\end{align}
        \label{eq:VAq}
	\end{subequations}
Combining Eq. \eqref{eq:VAq} we get the standalone expression for $n_0$ as $d_\xi^2 n_0 = -8 \alpha_q n_0$ that describes the oscillatory motion of the soliton with a period $\xi_c \approx  \pi/\sqrt{2\alpha_q}$. When the central waveguide is illuminated at an angle, the evolution of the beam's position \( n_0(\xi) \) and transverse wavevector \( \kappa_0(\xi) \) is described by \( n_0(\xi)= n_{max} \sin(2\sqrt{2\alpha_p}\xi) \) and \( \kappa_0(\xi)= \kappa_0(0)\cos(2\sqrt{2\alpha_q}\xi) \), respectively, where \( n_{max}= \kappa_0(0)/\sqrt{2\alpha_q} \). In contrast, when a soliton is launched normally and offset from the central waveguide, the evolution is given by \( n_0(\xi) = n_0(0) \cos(2 \sqrt{\alpha_q }\xi) \) and \( \kappa_0(\xi) = \kappa_{max} \sin(2 \sqrt{2 \alpha_q} \xi) \), with \( \kappa_{max} = - \sqrt{2 \alpha_q} n_0(0) \). The beam evolution in \( n \) and \( \kappa \) space is illustrated in \figref{fig:quad_ds} (a)-(d), showing amplitude and phase variation, which is supported by analytical results aligning well with numerical calculations. For a chirping strength of \( \delta=0.25 \) nm and \( \alpha_q=7.25 \times 10^{-5} \), the numerical period \( \xi_c \) closely matches the analytical prediction at approximately \( 260 \). Visual representations of the beam path for varying launching angles and initial positions are provided in plot (e),(f), confirming the alignment of numerical data with analytical predictions. 
{The effect of $\alpha_q$ on stationary soliton beam-width is further documented in \figref{fig:quad_ds} (g)-(h) revealing marginal compression.}
The study reveals that solitons follow a sinusoidal trajectory in $n$-space, demonstrating distinct variation patterns in the transverse wavenumber $\kappa_0$: zig-zag for linear chirping and sinusoidal for quadratic chirping. The proposed VA effectively models the soliton beam dynamics for both chirping schemes. Additionally, expressions for the maximum displacement of solitons in both $n$ and $\kappa$ space have been derived, which align closely with numerical simulations.

	\section{Conclusion}

The article explores the dynamics of beam evolution in waveguide arrays, which serve as a compact platform for various applications in solid-state and quantum physics. We propose two configurations of symmetrically chirped waveguide arrangements, with varying separations between adjacent channels following linear or quadratic functions. By applying a continuous approximation, we derive a coupling coefficient that varies with the transverse coordinate, allowing for a modified DNLSE that accounts for chirping effects. The validity of the derived equations is confirmed by comparison with the original DNLSE, showing consistent results. The continuous counterpart facilitates the use of semi-analytical variational techniques, leading to ordinary differential equations that describe the complex dynamics of beam propagation. Notably, the study finds that the behavior of a chirped waveguide array parallels that of a graded index medium, as the variable coupling coefficient plays a role similar to the refractive index in such systems.
We investigate the behavior of Gaussian beam propagation in a chirped waveguide array, focusing on the linear regime without Kerr nonlinearity. We establish a steady state solution analytically, noting that deviations from steady state lead to oscillatory propagation, consistent with variational predictions. In the presence of nonlinearity, however, the Gaussian beam fails to maintain a steady state, as it becomes an inaccurate solution. Despite this, numerical methods can yield stationary solutions, which are analyzed for stability. The study reveals that under nonlinearity, the emergence of discrete solitons is inevitable, with dynamics explored through sech-form solutions that balance discrete diffraction with nonlinearity. The results indicate unique oscillatory behaviors influenced by the perturbations of waveguide chirping, aligning with analytical forecasts. Ultimately, our research highlights the potential of optical lattices, particularly chirped waveguide arrays, for advanced control and routing of light, offering insights that could prove beneficial for future applications.

	\begin{acknowledgments}
		A.P.L. and S.R. are grateful to Anusandhan National Research Foundation (ANRF), erstwhile SERB, India, for providing financial support to carry out this work under the CRG program.
	\end{acknowledgments}

	\appendix

	\section{Derivation of the approximated continuous form of DNSE \label{appen1}}
	
	\subsection{Linear Chirping} \label{appen_lc}
	Considering the coupling profile for the symmetric linear chirping as $\eta_n^{n+1} \approx e^{-\alpha_l \abs{n+\frac{1}{2}}}$, we can evaluate the term $\eta_n^{n+1} \pm \eta_{n-1}^n$ present in Eq. \eqref{eq:cont_v1}.
	To proceed with this evaluation we have to consider the two domains of $n$, i.e $n \geq 0$ and $n < 0$, where $\eta_n^{n+1} = \exp\left[-\alpha_l (n + 1/2)\right]$ for $n \geq 1$, and $\eta_n^{n+1} = \exp \left[\alpha_{l}(n + 1/2)\right]$.
	\begin{equation}
		\eta_n^{n+1} \pm \eta_{n-1}^n =
		\begin{cases}
			e^{-\alpha_l n}\left(e^{-\alpha_l/2} \pm e^{\alpha_l/2} \right) & n \geq 0 \\
			e^{\alpha_l n} \left(e^{\alpha_l/2} \pm e^{-\alpha_l/2}\right) & n < 0
		\end{cases}
	\end{equation}
	Exploiting the identities $2\cosh(x) = e^x + e^{-x}$ and $2 \sinh(x) = e^x - e^{-x}$, we can write Eq. \eqref{eq:cont_v1} in the two regimes as, 
	\begin{align}
		i \partial_{\xi} \psi + \mathcal{N}^2\abs{\psi}^2 \psi + 2\cosh\left(\frac{\alpha_l}{2}\right)e^{-\alpha_l n}\left(\psi + \frac{1}{2} \partial_n^2 \psi\right) \nonumber \\
		- 2 \sinh\left(\frac{\alpha_l}{2}\right)e^{-\alpha_l n} \partial_n \psi = 0
	\end{align}
	for $n \geq 0$, and 
	\begin{align}
		i \partial_{\xi} \psi + \mathcal{N}^2\abs{\psi}^2 \psi + 2 \cosh\left(\frac{\alpha_l}{2}\right)e^{\alpha_l n} \left(\psi + \frac{1}{2}\partial_n^2 \psi\right) \nonumber \\
		+ 2 \sinh\left(\frac{\alpha_{l}}{2}\right)e^{\alpha_l n}\partial_n \psi = 0
	\end{align}
	for $n<0$. Using the signum function $\sgn(n)$ for $n$, the two equations can be written in a single equation as 
	\begin{align}
		i\partial_{\xi}\psi+\partial_{n}^{2}\psi+2\psi+ \mathcal{N}^2\left|\psi\right|^{2}\psi \nonumber \\  +\left[\cosh\left(\frac{\alpha_{l}}{2}\right)e^{-\alpha_{l}\left|n\right|}-1\right]\left(2\psi+\partial_{n}^{2}\psi\right) \nonumber \\ -2\sgn(n)\sinh\left(\frac{\alpha_{l}}{2}\right)e^{-\alpha_{l}\left|n\right|}\partial_{n}\psi=0.
		\label{eq:cont_linear_1}
	\end{align}
	In this expression, the first four terms are the standard terms of the NLSE, fifth and sixth term correspond the effective potential and variation in diffraction coefficient along the transverse axis due to the changing coupling coefficient.
	The last term is a transverse velocity imparted to the field due to the asymmetry in the coupling coefficient about the $n^\text{th}$ waveguide. Applying the Taylor series expansion of $\cosh(\alpha_l/2)$, $\sinh(\alpha_l/2)$, and $\exp(-\alpha_{l} \abs{n})$ and neglecting the terms containing higher order of $\alpha_l$ (as $\alpha_l \approx 10^{-3}$) we may write the approximate continuous NLSE as 
	\begin{align}
		i \partial_{\xi}\psi + \partial_{n} \psi + 2\psi + \mathcal{N}^2\abs{\psi}^2 \psi  - \alpha_l \abs{n}(2 \psi + \partial_n^2 \psi) \nonumber \\
		- \sgn(n) \alpha_l \partial_n \psi = 0
	\end{align}
	
	\subsection{Quadratic Chirping} \label{appen_qc}
	In the case of the quadratic chirping the functional form of the coupling coefficient profile is obtained as $\eta_n^{n+1} \approx \exp\left[-\alpha_q \left(n + \frac{1}{2}\right)^2\right]$.
	With this form, the $\eta_n^{n+1} \pm \eta_{n-1}^n$ terms can be expressed as 
	\begin{align}
		\eta_n^{n+1} \pm \eta_{n-1}^n = e^{-\alpha_q\left(n^2 + 1/4\right)}\left(e^{-\alpha_q n} \pm 
		e^{\alpha_q n}\right).
	\end{align}
	This modification leads to the continuous approximation of the DNLSE as 
	\begin{align}
		i\partial_{\xi}\psi+2\psi+\partial_{n}^{2}\psi+\mathcal{N}^2\left|\psi\right|^{2}\psi \nonumber \\
		+\left[\cosh\left(\alpha_{q}n\right)e^{-\alpha_{q}\left(n^2+1/4\right)}-1\right]\left(2\psi+\partial_{n}^{2}\psi\right) \nonumber \\
		-2\sinh\left(\alpha_{q}n\right)e^{-\alpha_{q}\left(n^{2}+1/4\right)}\partial_{n}\psi=0.
		\label{eq:cont_quad_1}
	\end{align}
	Expanding the functions involved in the equation in Taylor series and neglecting the higher order terms of $\alpha_q$ we further simplify the governing equation and obtain the expression as, 

	\begin{align}
		i \partial_{\xi}\psi + 2 \psi + \partial_n^2 \psi + \mathcal{N}^2\abs{\psi}^2 \psi - \alpha_q n^2 \left(2\psi + \partial_n^2 \psi\right) \nonumber \\
		- 2\alpha_q n\partial_n \psi =0
	\end{align}

	\bibliography{references.bib}

\end{document}